\documentclass[10pt,a4paper,aps,prl,twocolumn]{revtex4-1}

\usepackage{graphicx}
\usepackage[update,prepend]{epstopdf}
\usepackage{amsmath}
\usepackage{amssymb}
\usepackage{color}

\begin{document}

\title{Phonon-Driven Electron Scattering and Magnetothermoelectric Effect in Two-Dimensional Tin Selenide}

\author{Kaike Yang}
\author{Ji-Chang Ren}
\author{Hongfei Qiu}
\author{Jian-Sheng Wang}
\affiliation{Department of Physics, National University of Singapore, Singapore 117551, Republic of Singapore}

\date{\today}

\begin{abstract}
The bulk tin selenide (SnSe) is the best thermoelectric material currently with the highest figure-of-merit due to the strong phonon-phonon interactions. We investigate the effect of electron-phonon coupling (EPC) on the transport properties of two-dimensional (2D) SnSe sheet. We demonstrate that EPC plays a key role in the scattering rate where the constant relaxation time approximation is deficient. The EPC strength is especially large in contrast to that of pristine graphene. The scattering rate depends sensitively on the system temperatures and the carrier densities when the Fermi energy approaches the band edge. We also investigate the magnetothermoelectric effect of the 2D SnSe. It is found that at low temperatures there are enormous magnetoelectrical resistivity and magnetothermal resistivity above 500\%, suggesting the high potential for device applications. Our results agree reasonably well with the experimental data. 
\end{abstract}
\maketitle


The thermoelectric effect which enables direct conversion from waste heat to useful electricity offers a viable route for power generation and green energy development with no noise and free gas emissions \cite{Goldsmid2010}. The efficiency of the thermoelectric conversion is dominated by a dimensionless figure-of-merit \cite{Disalvo1999,Vining2008,Dubi2011,Yang2012,Dagosta2013}: $zT=\sigma S^2 T/(\kappa_e+\kappa_{ph})$, where $\sigma$ is the electrical conductivity, $S$ is the Seebeck coefficient, $\kappa_e$ and $\kappa_{ph}$ are the electron and phonon thermal conductivities, respectively, and $T$ is the absolute temperature. Obviously, an ideal thermoelectric material should hold the electrical conductivity and the Seebeck coefficient as high as possible but keep the thermal conductivity as low as possible. In the past several decades a variety of strategies including theory and experiment have emerged. For example, one expects by modifying the band structure \cite{Heremans2008} or by quantum confinement effect \cite{Hicks1993,Hicks1993a,Mahan1996} to enhance the Seebeck coefficient; On the other hand, to reduce the lattice thermal conductivity, nanostructuring the material is proposed \cite{Hochbaum2008,Boukai2008,Ni2009,Yang2014,Yang2015}. Nevertheless, the interdependence of $\sigma$, $S$ and $\kappa$ complicates the efforts for improving a material's $zT$ well above unity in a broaden range of temperatures.

Tin selenide (SnSe), a simple compound consisting of earth-abundant and nontoxic elements, can be derived from a distortion of the rocksalt structure \cite{Chattopadhyay1986a}. Recently it was surprisingly found that single-crystal SnSe exhibits a record high figure-of-merit $zT=2.6$ at $T=923$ K along a specific crystallographic direction \cite{Zhao2014,Zhao2016a,Ibrahim2017}, releasing a sign that it may cause a revolution in the field of thermal energy conversion. The particularly high figure-of-merit is attributed to the ultralow lattice thermal conductivity originating from the phonon anharmonic effect \cite{Zhao2014,Carrete2014,Skelton2016}. Li~et~al.~\cite{Li2015} reported that the giant anharmonicity in SnSe stems from the bonding instability which is determined by the long-range resonant Se $4p$-orbitals coupled to the Sn $5s$-orbitals. The orbital interaction between the Sn and Se atoms suggests the electron-phonon coupling (EPC) in this layered material may not be negligible. Clarification of how EPC influences the transport characteristics of SnSe could not only deepen the understanding of the interacting physical picture but also benefit us when designing novel devices in future for thermoelectric applications.

In this Letter, we investigate the thermoelectric transport properties of two-dimensional (2D) SnSe sheet by taking into account the phonon scatterings. We employ the deformation potential theory \cite{Bardeen1950,Herring1956,Gantmakher1987} to calculate the EPC $g$-factor and then the electron scattering rates with respect to different phonon modes. It is found that the EPC strength in this 2D material is incredibly large in contrast to that of the pristine or weakly-doped graphene \cite{Ulstrup2012}. The total scattering rate is mainly contributed by the longitudinal acoustic and optical phonon modes, while the shear modes' contributions compared to the former are almost negligible. Through tuning the working temperatures and the carrier densities, we find that the scattering rate varies intensively when the Fermi energy is placed near the valence-band edges, indicating the deficiency of the constant relaxation time approximation. Based on the Boltzmann transport equations combined with the density-functional theory (DFT) calculations, we further investigate the thermoelectric coefficients in the presence of magnetic field which is perpendicular to the 2D SnSe plane [see Fig.~1(a)]. It is shown that there are enormous magnetoelectrical resistivity and magnetothermal resistivity above 500\% at low temperatures, suggesting the high potential for device applications in magnetic memory. Finally, we compare the theoretical results with the experiments \cite{Zhao2014,Zhao2016a} and find that they are in reasonably good agreement.

\paragraph{Band structures.}

It is well-known that bulk SnSe is an orthorhombic crystal which has two phases, i.e., the low temperature {\it Pnma} phase and the high temperature {\it Cmcm} phase, where the structure conversion via displacive atomic coordinates occurs at $T_C\approx 750$ K \cite{Zhao2014}. In this work we are interested merely in the phase of $T<T_C$ and the properties of SnSe near and above the phase transition temperature were discussed systematically in Refs.~[\onlinecite{Li2015,Dewandre2016a,Bansal2016a,Hong2016a}]. Our attentions are concentrated on the 2D SnSe sheet which was successfully synthesized through a so-called one-pot method recently \cite{Li2013a}. Figure~1(a) shows a top view of the single-layer SnSe sheet. Different from graphene, the tin and selenium atoms are not placed in the same plane. Instead, the 2D SnSe is constructed by double atomic sub-layers with interlayer distance $h_0=2.75$ {\AA} in the $z$ direction perpendicular to the 2D plane [see clearly from Fig.~S1 of the supporting information (SI)]. Along the $x$ (horizontal) direction, the adjacent atoms are bonded with a zigzag shape; While along the $y$ (vertical) axis, atoms are connected like an armchair edge [see Fig.~1(a)].

Before performing the electronic energy band calculations, we relax the structure of the 2D SnSe initially by using first-principle approach \cite{Kohn1965a} as implemented in {\it Quantum Espresso} package \cite{Giannozzi2009}. The generalized-gradient approximation (GGA) which was parametrized in the form of Perdew-Burke-Ernzerhof (PBE) \cite{Perdew1996a} for the exchange-correlation potential, together with the projector-augmented wave (PAW) method \cite{Blochl1994}, is adopted. For the self-consistent potential and total energy calculations, the k-points of the first Brillouin-zone are sampled by a $20\times 20$ mesh \cite{Monkhorst1976}. As to the non-self-consistent field calculations, we sample the k-points by a $150\times 150$ mesh to obtain the energy bands and to achieve the converged results in computation of the thermoelectric transport coefficients (see below). The kinetic energy cut-off for the plane wave basis set is chosen as 500 eV \cite{Kresse1996a}. The convergence threshold of energy is set to be $10^{-8}$ eV and the convergence threshold of force acting on each atom is $10^{-4}$ eV/{\AA}. To avoid spurious interactions between the replica of the 2D SnSe sheet introduced by the periodic boundary condition, we have included a vacuum gap larger than 15 {\AA} in the out-of-plane, i.e., $z$-direction of the unit cell. After relaxation, we obtain the lattice constants of the single-layer SnSe as $a_0=4.30$ {\AA} and $b_0=4.37$ {\AA} for the $x$ and the $y$ crystallographic directions, respectively. Our optimized structure parameters are in excellent agreement with the previous report \cite{Wu2016a}.

\begin{figure}[t!]
\includegraphics[width=8.5cm]{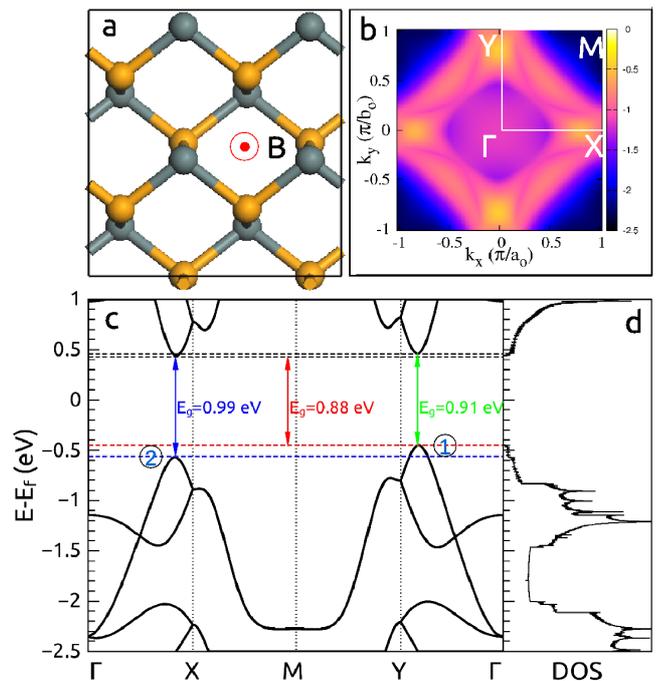}
\caption{(Color online) (a) Schematic illustration of a top view of the single-layer SnSe sheet with ball-stick representation, where an enlarged $2\times 2$ supercell is presented, the gray and the yellow atoms can stand for either tin or selenium elements, and the magnetic field $B$ for the study of magnetothermoelectric properties (see below) is perpendicular to the principal plane; (b) DFT calculated band structure of the full irreducible first Brillouin-zone and the high symmetric points are labelled. We use $a_0$ and $b_0$ to denote the lattice constants of the 2D SnSe in the $x$ ($\Gamma\rightarrow$X) and $y$ ($\Gamma\rightarrow$Y) directions, respectively; (c) Energy band structure of the 2D SnSe with respect to the high-symmetric lines shown in (b), where the Fermi energy $E_f$ is set as 0, $\textcircled{1}$ and $\textcircled{2}$ denote the first and second valence bands for convenience in discussions. We have also denoted the direct and indirect band gaps for a comparison with the experiment (see main text); (d) Density of states plotted as a function of electron energy which has the same scale of (c).}
\label{2Dbands}
\end{figure}

Once we obtained the optimized geometrical structure of the 2D SnSe, we could calculate the electronic energy bands. Figure~1(b) shows a contour plot of the bands of the 2D SnSe in the full irreducible first Brillouin-zone, where the high symmetric points connected by solid lines are depicted and will be used for the following calculations. We see clearly that there are four vertices with the highest energy of the valence bands and each band vertex closed to the Fermi energy is located in the vicinity of either X ($-$X) or Y ($-$Y) points. Electrons near the X and Y points will contribute principally to the transport (see below) since the other states are too far away from the Fermi energy. A more intuitive result is shown in Fig.~1(c), which plots the energy band as a function of the high symmetric line shown in (b). It is found that the first two valence bands $\textcircled{1}$ and $\textcircled{2}$ along the $x$ and $y$ directions are not precisely the same, indicating an anisotropic characteristic. The electronic energy at the top of band $\textcircled{2}$ is slightly lower than that of the band $\textcircled{1}$, which suggests that at low temperatures the states belong to band $\textcircled{1}$ dominate the transport behavior. By checking the conduction bands, we find that the direct band gaps near the X and Y positions are $E_g=0.99$ eV and $E_g=0.91$ eV, respectively. The indirect band gap, however, for this material corresponds to 0.88 eV. Our band structure calculations are consistent with the previous theoretical \cite{Gomes2015,Guo2015,Ding2015a,Wu2016a} and experimental results \cite{Li2013a,Zhao2014,Li2015,Li2017} since we are aware that for this single-layer SnSe, the measured indirect and direct band gaps are 0.86 eV and 1.1 eV, respectively. The good agreement in band gap between GGA-PBE and experiment to some extent is due to the PAW potentials which capture the six valence electrons of Se ($4s^2 4p^4$) and the four of Sn ($5s^2 5p^2$) properly.

From the band structures of SnSe we extract the density of states (DOS) as shown in Fig.~1(d), which presents unambiguously the band gap range. Furthermore we find that there is an obvious step at the maximum occupied states, indicating a nearly linear energy dispersion in the valance bands. By analyzing the projective density of states (PDOS) for each component of SnSe [see Fig.~S2 of SI], we notice that the valence bands $\textcircled{1}$ and $\textcircled{2}$ are mainly determined by the Se atoms with the in-plane $p$-orbitals, while the Sn components are responsible to the low energy regime of the conduction bands. The behavior can be easily understood from the chemical point of view since Se is a donor of electrons and Sn is an acceptor for SnSe.

\paragraph{Scattering rate with electron-phonon coupling.}

Based on the DFT band structure calculations, we concentrate here on the scattering effect of the single-layer SnSe sheet. The electron scattering rate due to EPC is determined by \cite{Mahan2000,Hwang2008,Kaasbjerg2012,Zhao2016,Liu2017}
\begin{equation}
\begin{split}
\frac{1}{\tau_{\mathbf{k},\Lambda}}=&\frac{\Omega}{2\pi h\hbar} \int_{BZ} d^2\mathbf{k'}~|g_{\mathbf{k},\mathbf{q}\Lambda}|^2 \left[N_{\mathbf{q}\Lambda}\delta(\varepsilon_\mathbf{k'}-\varepsilon_\mathbf{k}-\hbar\omega_{\mathbf{q}\Lambda})+\right. \\ &\left(1+N_{\mathbf{q}\Lambda})\delta(\varepsilon_\mathbf{k'}-\varepsilon_\mathbf{k}+\hbar\omega_{\mathbf{q}\Lambda})\right] \frac{1-f(\varepsilon_\mathbf{k'})}{1-f(\varepsilon_\mathbf{k})}\times \\
&\left(1-{\rm cos}{\rm \theta}_{\mathbf{k}\mathbf{k'}}\right)^{S_\Lambda}
\end{split}
\label{ScatteringRate_kspace}
\end{equation}
where the integral takes over the whole first Brillouin-zone ({\it BZ}); $\mathbf{k}$ (or $\mathbf{k'}$) and $\mathbf{q}$ are the electron and phonon momentum vectors, $\varepsilon_\mathbf{k}$ and $\omega_{\mathbf{q}\Lambda}$ are the electron band energy and the phonon vibrational frequency at a specific mode, respectively; $N_{\mathbf{q}\Lambda}=1/(e^{\hbar\omega_{\mathbf{q}\Lambda}/k_B T}-1)$ and $f(\varepsilon_\mathbf{k})=1/(e^{(\varepsilon_\mathbf{k}-\mu)/k_B T}+1)$ are the Bose-Einstein and Fermi-Dirac distribution functions at equilibrium states for a finite system temperature; $\hbar$ is the reduced Planck constant, $h=h_0+\delta$ with $\delta=2.73$ {\AA} for the consideration of atomic bonding in the $z$ direction, and $\Omega$ is the volume of the SnSe unit cell; $\mu$ and $k_B$ are the chemical potential and the Boltzmann constant, respectively; ${\rm \theta}_{\mathbf{k}\mathbf{k'}}$ describes the scattering angle between vectors $\mathbf{k}$ and $\mathbf{k'}$; $\Lambda$ corresponds to the different phonon branches, while the parameter ${S_\Lambda}=1$ if $\Lambda$ corresponds to the acoustic phonon branches and $0$ if $\Lambda$ for optical branches according to Refs.~[\onlinecite{Mahan2000},~\onlinecite{Nag1980}] due to the inelastic feature.

To calculate the electron scattering rate, we must know the EPC strength $g_{\mathbf{k},\mathbf{q}\Lambda}=\sqrt{\frac{\hbar}{2\rho\Omega\omega_{\mathbf{q}\Lambda}}} M_{\mathbf{k},\mathbf{q}\Lambda}$ in Eq.~(\ref{ScatteringRate_kspace}), where the transition matrix is defined as $M_{\mathbf{k},\mathbf{q}\Lambda}=\mathbf{e}_{\Lambda}(\mathbf{q}) \cdot\left<\mathbf{k}+\mathbf{q}\left|\frac{\partial H_e}{\partial \mathbf{R}}\right|\mathbf{k}\right>$, $\rho$ is the mass density, $\mathbf{e}_\Lambda(\mathbf{q})$ is the polarization vector for phonon branch $\Lambda$, $H_e$ is the electronic Hamiltonian and $\mathbf{R}$ is the atomic coordinate in the unit cell of SnSe.

We employ the deformation potential theory to calculate the EPC transition matrix $M_{\mathbf{k},\mathbf{q}\Lambda}$. This method is broadly used in the past for material's performance prediction and gives an extremely good fit of the experimental results \cite{Bardeen1950,Herring1956,Gantmakher1987,Attaccalite2010}. We take into account the phonon modes only at the $\Gamma$ point since these modes contribute mostly compared to the others in calculating the transport coefficients. For the acoustic and optical phonon modes, we consider mainly the longitudinal vibrational states, while the transverse mode's contribution according to Ref.~[\onlinecite{Gantmakher1987}] is much smaller than that of the longitudinal ones. Therefore, the longitudinal acoustic deformation potential $\Xi$ is determined by $M_{\alpha}=\Xi_{\alpha} \triangledown_{\alpha} {u_\alpha}$, where $\alpha=x$ or $y$ corresponding to the crystallographic direction of the 2D SnSe and ${u_\alpha}$ is the atomic displacement. We calculate $\Xi_x$ and $\Xi_y$ by stretching the lattice of the 2D SnSe along the $x$ and $y$ directions, respectively. For the optical phonon modes, we define $M_{\alpha}={\mathcal D}_{\alpha} {u_\alpha}$, where ${\mathcal D}_{\alpha}$ is the optical deformation potential. We calculate ${\mathcal D}_{\alpha}$ by shifting the Sn or Se atomic coordinates slightly in the unit cell. As to the shear modes, we define $M_{xy}=\frac{1}{2}\Upsilon_{xy}(\triangledown_x {u_y}+\triangledown_y {u_x})$ for the acoustic shear deformation potential and then $\Upsilon_{xy}$ is obtained by shearing the unit cell of the 2D SnSe. While for the optical shear deformation potential, it is defined by $M_{xy}=\frac{1}{2}{\mathcal W}_{xy}({u}_{xx}+{u}_{yy})$. We calculate ${\mathcal W}_{xy}$ by moving the Sn and Se coordinates of the upper and lower sub-layers of the 2D SnSe in an opposite direction. The distortion of the lattice creates a band shift which is essential of the theory to calculate the deformation potentials, and we present all the details for the practical implementation in the supporting information [see Fig.~S3].

\begin{figure*}[t!]
\includegraphics[width=18cm]{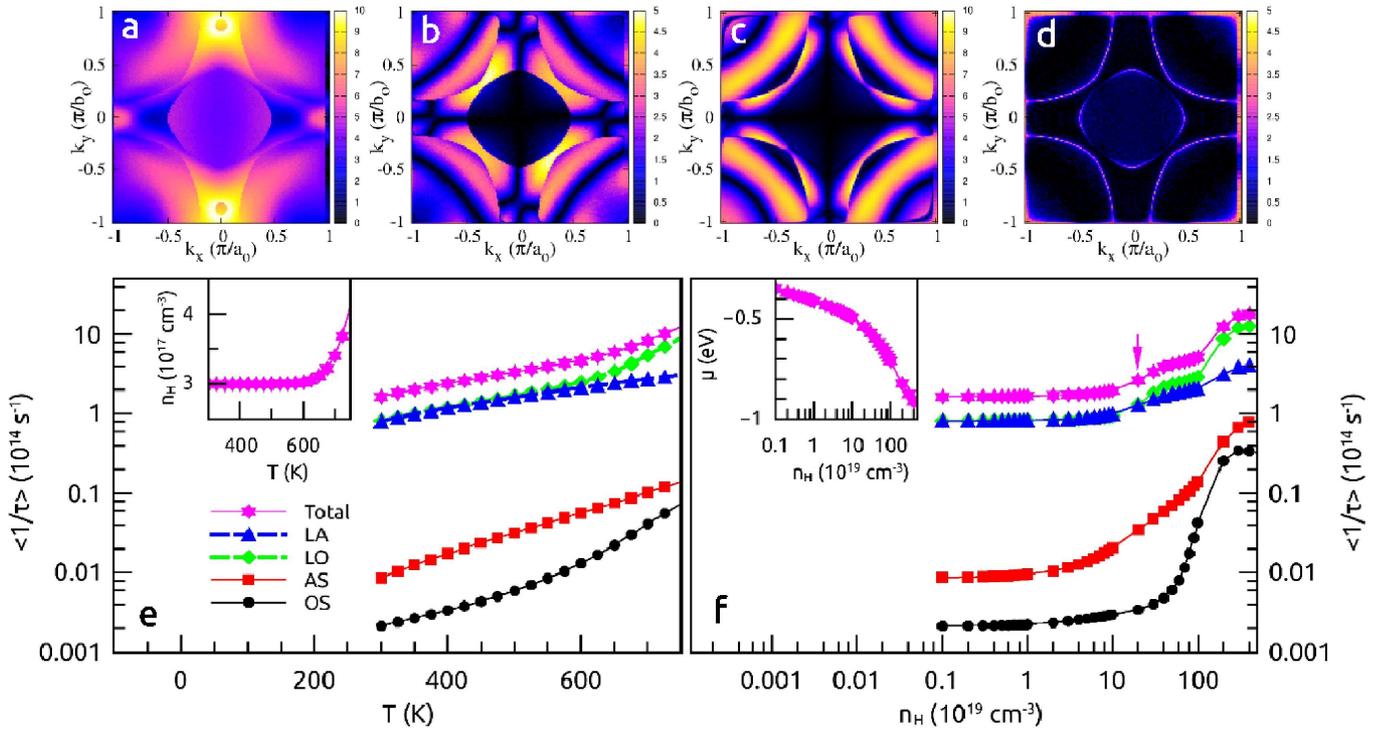}
\caption{(Color online) (a-d) $\mathbf{k}$-dependence of the deformational potentials in the first Brillouin-zone for different phonon branches: (a) $\Xi_x$, (b) ${\mathcal D}_x$, (c) $\Upsilon_{xy}$ and (d) ${\mathcal W}_{xy}$, where $\Xi_x$ and $\Upsilon_{xy}$ are in unit of eV, ${\mathcal D}_x$ and ${\mathcal W}_{xy}$ in unit of eV/{\AA}; (e) and (f) the scattering rates defined by Eq.~(\ref{ScatteringRate_average}) by taking the integrations over the conducting electrons for the single-layer SnSe, where the dashed blue and green lines in (e) are the polynomial fitting (see main text) and the fitted parameters are listed in Table~\ref{LA_LO_coefficients}, the arrow in (f) corresponds to the vertices of the valance bands $\textcircled{1}$ and $\textcircled{2}$. The inset in (e) shows the carrier density as a function of system temperature for a given particle number $\mathcal N$ (see main text), and the inset of (f) shows the chemical potential as a function of carrier density for temperature $T=300$ K.}
\label{ScatteringRate}
\end{figure*}

Figures~2(a-d) show the $\mathbf{k}$-dependent deformation potentials of the 2D SnSe sheet in the first Brillouin-zone for different phonon branches. We see that the longitudinal acoustic (LA) deformation potential $\Xi_x$ presents two maximums at the vertices of the valence band $\textcircled{1}$ near the Y and $-$Y points [see Fig.~1(c)]. As to $\Xi_y$, it is also found that there are two maximums but located near the X and $-$X points [see Fig.~S4(a) in SI]. Regarding to the longitudinal optical (LO) deformation potential shown in Fig.~2(b), the value compared to the acoustic one at the valence band edges is relatively small. By checking ${\mathcal D}_x$ and ${\mathcal D}_y$, we find that the contour of the $\mathbf{k}$-dependent optical deformation potentials are quite different [see Fig.~S4(b)], which is originated from the polarization nature of the optical phonons. In Figs.~2(c) and (d) we plot the deformational potentials corresponding to the acoustic and optical shear (AS and OS) branches. It is shown that the shear deformation potentials in contrast to the longitudinal $\Xi_x$ and $\Xi_y$ at the valence band edges X ($-$X) and Y ($-$Y) are much small, in particular for the OS branch. We shall see in the following that the shear mode's contribution for transport is of little significance and can be safely neglected.

Using the deformation potentials, we calculate the scattering rates with Eq.~(\ref{ScatteringRate_kspace}) for different phonon branches. To give a more intuitive picture of how EPC influences the transport behavior of SnSe, we introduce an average of the scattering rate by integrating over the conducting electrons and so we define
\begin{equation}
\left\langle\frac{1}{\tau_{\Lambda}}\right\rangle=\int_{BZ} \frac{\partial f(\varepsilon_\mathbf{k})} {\partial\varepsilon_\mathbf{k}} \frac{1}{\tau_{\mathbf{k},\Lambda}} d^2\mathbf{k} \bigg/\int_{BZ} \frac{\partial f(\varepsilon_\mathbf{k})}{\partial\varepsilon_\mathbf{k}} d^2\mathbf{k}
\label{ScatteringRate_average}
\end{equation}
where the denominator is a normalization factor and the partial derivative of the Fermi-Dirac function guarantees that only the states closed to the Fermi energy participate in the integral. Therefore, the total scattering rate
\begin{equation}
\left\langle\frac{1}{\tau_{\rm tot}}\right\rangle=\left\langle\frac{1}{\tau_{\rm LA}}\right\rangle+\left\langle\frac{1}{\tau_{\rm LO}}\right\rangle+\left\langle\frac{1}{\tau_{\rm AS}}\right\rangle+\left\langle\frac{1}{\tau_{\rm OS}}\right\rangle
\end{equation}
by summing over all the phonon branches that we are interested in.

Figure~2(e) shows the electron scattering rate in unit of $10^{14}$ per second as a function of system temperature for different phonon branches of the single-layer SnSe sheet. In calculations we have kept the particle number ${\mathcal N}$ as a constant and then change the temperatures. Accordingly the carrier density is modulated by resetting the Fermi level [see the inset]. We define ${\mathcal N}={\mathcal N}_{crys}-\Delta=\int_{-\infty}^{\infty} d\varepsilon f(\varepsilon) {\rm DOS}(\varepsilon) \Omega$, where ${\mathcal N}_{crys}$ is the fully occupied particle number of the perfect crystal at zero temperature, and $\Delta$ is a small quantity which is added by taking into account the fact that in practice it is unavoidably to have some defects such as vacancies in the sample. We set $\Delta=3\times 10^{17}{\rm cm}^{-3} \Omega$ to match with the experimental measurements [see Fig.~S6]. In this work, we focus only on the hole transport while the electron transport for conduction bands is not concerned. Therefore, the hole carrier density is determined by $n_{\rm H}=\int_{-\infty}^{\infty} d\varepsilon [1-f(\varepsilon)] {\rm DOS}(\varepsilon)/\Omega$. From Fig.~2(e) we find that the scattering rates for all the phonon branches increase with temperature. The scattering rates for LA and LO are much higher than that of the shear branches, demonstrating the case that the latter's contribution to transport compared to the former is negligible as we have mentioned before. At low temperatures, it is shown that the LA and LO branches contribute almost equally to the transport. At high temperatures, however, the scattering rate for LO becomes more significant.

To see more clearly the temperature-dependence behavior of the scattering rates, we use a polynomial to fit the data [see the dashed lines of Fig.~2(e)]. For the LA phonon branch, it is found that we must use $\left\langle\frac{1}{\tau_{\rm LA}}\right\rangle= \frac{1}{\tau_0}\sum_{i=0}^{p}c_i T^i$ up to $p=2$ to get well fitted, where $\tau_0=10^{-14}$ second and $c_i$ is the fitted parameter. At very low temperatures, the scattering rate can be regarded as a constant. With increasing the temperature as $T>225$ K, the linear temperature-dependence of the scattering rate becomes important. When $T>500$ K, the quadratic term starts to contribute. For the LO branch, it is shown that we have to use the polynomial until $p=4$ to obtain the excellent agreement with the calculations and the fitted parameters are listed in Table~\ref{LA_LO_coefficients}. The third-order and fourth-order, nevertheless, become significant only when $T>600$ K. The nonlinear dependence of $1/\tau$ on $T$ should date back to Eq.~(\ref{ScatteringRate_kspace}) with the Fermi-Dirac distribution function.

\begin{table}[b!]
\caption{Polynomial fitted parameters for the scattering rates with respect to the LA and LO phonon branches.}
\begin{ruledtabular}
\begin{tabular}{cccccc}
 & $c_0$ & $c_1~(/{\rm K})$ & $c_2~(/{\rm K}^2)$ & $c_3~(/{\rm K}^3)$ & $c_4~(/{\rm K}^4)$ \\
\hline
 LA & 0.19 & 8.53$\times$ $10^{-4}$ & 4.02$\times$ $10^{-6}$ & & \\
\hline
 LO & 10.11 & -0.113 & 4.7$\times$ $10^{-4}$ & -8.1$\times$ $10^{-7}$ & 5.1$\times$ $10^{-10}$ \\
\end{tabular}
\end{ruledtabular}
\label{LA_LO_coefficients}
\end{table}

In Fig.~2(f) we plot the scattering rate versus the hole carrier density, where we have set the system temperature $T=300$ K. The carrier density is given according to the Fermi level or the chemical potential as shown in the inset and then the scattering rate is calculated. It is found that $\left\langle\frac{1}{\tau}\right\rangle$ for all the phonon branches at $n_{\rm H}<10^{19}{\rm cm}^{-3}$ increases very slowly. This is because the chemical potential is too far away from the valence band edges. With increasing the carrier density, the scattering rate undergoes a tremendous enhancement and the transition happens at $n_{\rm H}\approx 2\times 10^{20}{\rm cm}^{-3}$ (see the arrow), which corresponds to $\mu$ at the vertices of the valence bands $\textcircled{1}$ and $\textcircled{2}$ [see Fig.~1(c)]. With further increasing the carrier density, we find again a rapid enhancement of the scattering rate which is due to more and more valence bands participating in the transport. From the results we acquire that one has to include the energy-dependence in computation of the relaxation time when the Fermi level is close to the band edges. The intensively doping-level dependence of the scattering rates indicates that the constant relaxation time approximation in this 2D material is not valid.

\begin{figure}[t!]
\includegraphics[width=8.5cm]{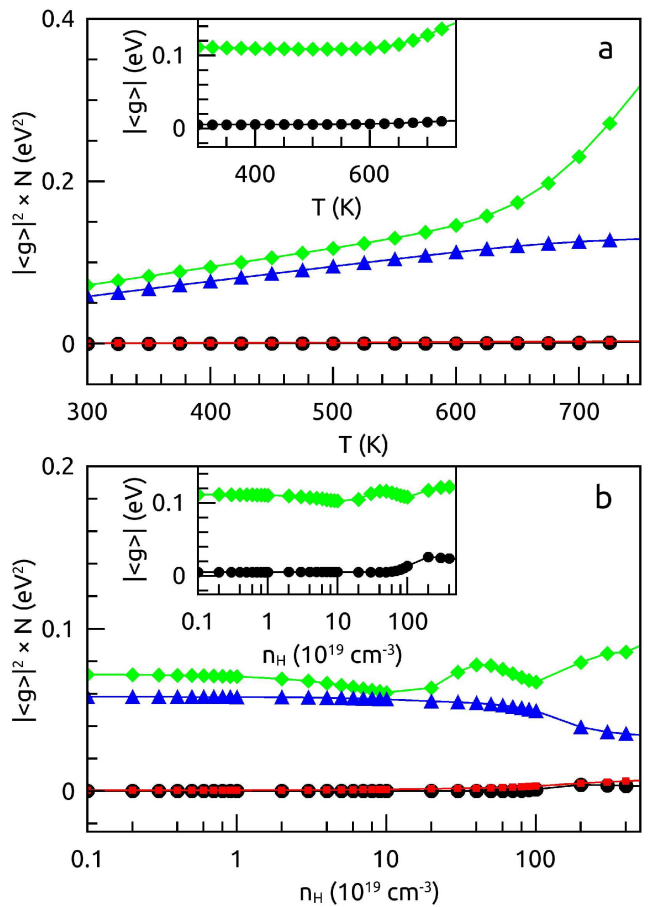}
\caption{(Color online) The EPC strength as a function of (a) system temperature for a given particle number ${\mathcal N}$ and (b) carrier density for temperature $T=300$ K, where the equilibrium Bose-Einstein distribution function $N$ is multiplied to avoid the singularity of the $g$-factor for acoustic modes. The insets in (a) and (b) show the EPC strength of the LO and LA branches. The symbols (circle, square, diamond and triangle) with respect to different phonon branches have the same meaning of Fig.~2(e).}
\label{EPCoupling}
\end{figure}

The origin of the increase of the scattering rate with system temperatures can be understood from the EPC strength. In Fig.~3(a) we plot $|\langle g\rangle|^2\times N$ as a function of temperature, where $\langle g\rangle$ is calculated according to Eq.~(\ref{ScatteringRate_average}) by simply replacing $\frac{1}{\tau_{\Lambda}}$ with $g_{\Lambda}$. The equilibrium Bose-Einstein distribution function $N$ is multiplied to avoid the singularity of the $g$-factor for LA and AS branches since we consider phonons at the $\Gamma$ point. In the inset of Fig.~3(a) we plot $\langle g\rangle$ for LO and OS branches to see directly the coupling strength. It is found that the EPC strength for AS and OS is extremely small, approaching zero. At low temperatures, the coupling strength for LO and LA is quite close to each other with $\langle g\rangle\approx 0.11$ eV. Nevertheless, with increasing the temperature as $T>600$ K, the coupling strength for LO increases quickly and is much higher than the value of LA branch, confirming that at high temperatures the optical modes play dominant role in the scattering. Figure~3(b) plots the EPC strength of the 2D SnSe as a function of carrier density. Again we observe that the shear modes' contributions are small enough. When the Fermi level is far away from the band edges, the EPC strength is almost a constant for both LA and LO branches. As $\mu$ closing to the valence band edges, the $\langle g\rangle$-factor varies slightly with the carrier density [see the inset of Fig.~3(b)].

The obtained EPC strength of the 2D SnSe compared to that of the pristine or weakly-doped graphene is quite large, since the latter according to Ref.~[\onlinecite{Ulstrup2012}] has a very small number ($<0.001$) at the Dirac cone \cite{reference2017}. The large EPC value is originated from the strong lattice vibration coupled to the electronic states and can be understood from the band shift of the distorted SnSe compared to the perfect crystal. In Figs.~S5(a) and (b) we plot the band structures of the 2D SnSe by stressing the lattice along the $x$ and the $y$ directions, respectively. It is found that at the vertices of the valence bands $\textcircled{1}$ and $\textcircled{2}$, an obvious energy difference is presented.

It should be mentioned that in the above calculations we have not included the impurity or defect scattering which may be significant in a heavy doping semiconductor material. The system we considered here, however, is a single-crystal 2D SnSe and we expect that the impurity scattering is of little importance compared to EPC. Our transport analyses confirm the case since the modelling thermoelectric coefficients agree reasonably well with the experimental data [see Fig.~S6 of SI]. This research provides a simple and an efficient way to study the electron-phonon scattering.

\paragraph{Magnetothermoelectric effect.}

The calculation of the electron scattering rate motivates us to study the thermoelectric properties of the 2D SnSe sheet. We utilize the Boltzmann transport equations to calculate the electrical and thermal conductivities as well as the Seebeck coefficient. The Lorentz function in the presence of magnetic field according to Ref.~[\onlinecite{Ziman1960}] is given by
\begin{subequations}
\begin{equation}
\begin{split}
{\mathcal L}_{n,\alpha\alpha}=&\int_{BZ} \frac{d^2\mathbf{k}}{2h\pi^2}\left(-\frac{\partial f(\varepsilon_\mathbf{k})}{\partial\varepsilon_{\mathbf{k}}}\right) \tau_{\mathbf{k}} (\varepsilon_\mathbf{k}-\mu)^n \times \\ &\left[v_{\mathbf{k}_\alpha}v_{\mathbf{k}_\alpha}\Pi_{\mathbf{k}\alpha} -s_\alpha v_{\mathbf{k}_\alpha}v_{\mathbf{k}_\beta}\Pi_{\mathbf{k}\beta}e\tau_{\mathbf{k}}B/m_{\mathbf{k}\beta}\right]
\end{split}
\label{Lorenznumberxx}
\end{equation}
\begin{equation}
\begin{split}
{\mathcal L}_{n,\alpha\beta}=&\int_{BZ} \frac{d^2\mathbf{k}}{2h\pi^2}\left(-\frac{\partial f(\varepsilon_\mathbf{k})}{\partial\varepsilon_\mathbf{k}}\right) \tau_\mathbf{k} (\varepsilon_\mathbf{k}-\mu)^n \times \\ &\left[v_{\mathbf{k}_\alpha}v_{\mathbf{k}_\beta}\Pi_{\mathbf{k}\beta} +s_\alpha v_{\mathbf{k}_\alpha}v_{\mathbf{k}_\alpha}\Pi_{\mathbf{k}\alpha}e\tau_\mathbf{k}B/m_{\mathbf{k}\alpha}\right]
\end{split}
\label{Lorenznumberyy}
\end{equation}
\end{subequations}
where $\Pi_{\mathbf{k}\alpha}=1/[1+\left(e\tau_{\mathbf{k}}B/m_{\mathbf{k}\alpha}\right)^2]$ and $\Pi_{\mathbf{k}\beta}=1/[1+\left(e\tau_{\mathbf{k}}B/m_{\mathbf{k}\beta}\right)^2]$; $\tau_\mathbf{k}$ is the total relaxation time including all the phonon branches considered above and we have omitted the mode index for simplicity, $B$ is the magnetic field which has the direction of $z$ axis [see Fig.~1(a)], $\alpha=x$ and $\beta=y$ or reversely; $v_{\mathbf{k}_\alpha}$ is the group velocity which is defined according to $v_{\mathbf{k}_\alpha}=\frac{\partial \varepsilon_\mathbf{k}}{\partial {\hbar\mathbf{k}_\alpha}}$, $m_{\mathbf{k}\alpha}$ is the electron effective mass determined by $m_{\mathbf{k}\alpha}={\hbar^2}\big/{\frac{\partial^2 \varepsilon_\mathbf{k}}{\partial {\mathbf{k}^2_\alpha}}}$, similarly for $v_{\mathbf{k}_\beta}$ and $m_{\mathbf{k}\beta}$; the sign function $s_\alpha=1$ if $\alpha=x$ and $s_\alpha=-1$ if $\alpha=y$. The equation can be derived formally if one takes that the Lorentz force deviates only the electron path but not produces the drift on the charge.

Within linear response theory, using the Onsager relations \cite{Onsager1931,Onsager1931a}, the electrical conductivity can be expressed in terms of the Lorentz coefficient as $\sigma=e^2 {\mathcal L}_0$, where $e$ is the magnitude of electron charge. The electric thermal conductivity is therefore determined by $\kappa_e=\frac{1}{T}\left({\mathcal L}_2-{{\mathcal L}_0}^{-1}{{\mathcal L}^2_1}\right)$. Meanwhile, we have also defined the electrical resistivity $\rho={\sigma}^{-1}$ and the thermal resistivity $r={\kappa_e}^{-1}$ for the discussion of the magnetothermoelectric conversion (see below). As to the Seebeck coefficient, it is given by $S=-\frac{1}{eT}{{\mathcal L}_0}^{-1}{{\mathcal L}_1}$ \cite{Jiang2011,Lu2016,Yang2016b}.

The validity of our transport calculations is checked by making a comparison with the experiments for the thermoelectric coefficients of the 2D SnSe in the absence of magnetic field. The results are shown in the supporting information. Because of the lacking of the experiment in 2D SnSe, we adopt the data of the corresponding bulk material from Refs.~[\onlinecite{Zhao2014},~\onlinecite{Zhao2016a}] as a reference. Figures~S6(a-d) show the electrical conductivity, the electric thermal conductivity, the Seebeck coefficient and the electronic figure-of-merit $zT_e$, respectively, where $zT_e=\sigma S^2 r T$ by neglecting the phonon conduction. We have calculated from $n=1$ to $n=4$, where $n$ is the number of the atomic-layers of the 2D SnSe. We have set the particle number $\mathcal N$ as a constant as the same of Fig.~2(e). It can be seen that all the thermoelectric coefficients at $T<500$ K agree reasonably well with the experimental data, although the magnitudes are not precisely the same. The discrepancy between our theory and the experiments should be expected. With increasing the number of atomic-layers, it is noticed that the thermoelectric coefficients for $n=1,2,4$ behave similarly with the increasing of $T$. In Figs.~S6(e-h) we discuss the case of the doped 2D SnSe and the carrier density $n_{\rm H}=4\times 10^{19}{\rm cm}^{-3}$ to be consistent with the experiments. Again we find that our calculations are in agreement with the measurements.

\begin{figure*}[t!]
\includegraphics[width=16cm]{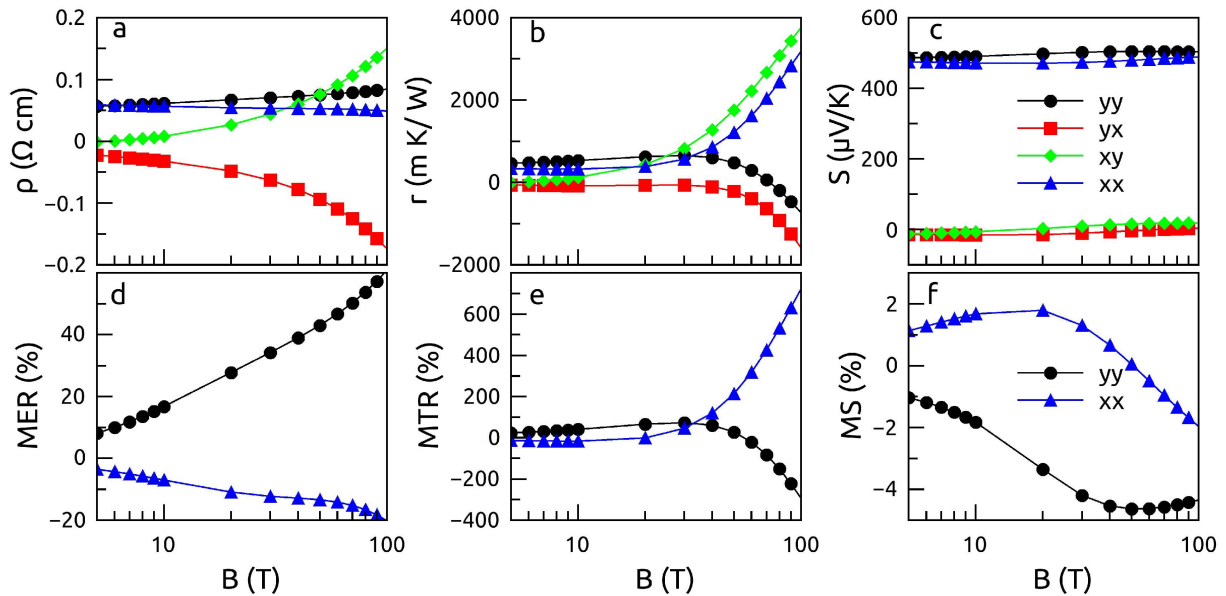}
\caption{(Color online) (a) electrical resistivity, (b) thermal resistivity and (c) Seebeck coefficient as a function of magnetic field $B$ in unit of Tesla, where the different colors represent the diagonal ($xx$ in blue and $yy$ in black) and off-diagonal ($xy$ in green and $yx$ in red) components of the corresponding quantities; (d-f) are the magnetoelectrical resistivity, magnetothermal resistivity and magnetoseebeck coefficient of the single-layer SnSe, respectively. We have set the particle number ${\mathcal N}$ as a constant (see main text) and the system temperature $T=100$ K in the calculations.}
\label{MTE_coefficients}
\end{figure*}

Finally we analyse the magnetothermoelectric properties of the 2D SnSe. Figures~4(a-c) show the electrical resistivity $\rho$, the thermal resistivity $r$ and the Seebeck coefficient $S$, respectively, where the various colors represent different components of the corresponding physical quantities. In calculations we have set the system temperature $T=100$ K and the other parameters are the same as Fig.~2(e). It can be seen that the diagonal $xx$ and $yy$ components of the electrical resistivity change slowly with $B$, while the off-diagonal Hall components increase quickly. The negative value in $\rho_{yx}$ indicates that the charge transport along the opposite direction of our defined crystal axis. For the thermal resistivity, we find that $r$ at small $B$ changes slowly. Nevertheless, when $B>30$ Tesla, it behaves almost linearly [see Fig.~4(b)]. In Fig.~4(c) we plot the Seebeck coefficient versus the magnetic field. It is shown that the Seebeck coefficient is much less sensitive to $B$ even at high magnetic field regime. At low magnetic field, since $\left(e\tau B/m\right)$ is almost zero, the second term in Eqs.~(\ref{Lorenznumberxx},~\ref{Lorenznumberyy}) can be ignored. Indeed, we observe the Hall resistivities $\rho_{xy}$ and $\rho_{yx}$ are almost zero and $\rho_{xx}$ and $\rho_{yy}$ are nearly constants. Similar happens for $r$. When $B$ is large enough, $\left(e\tau B/m\right)\gg 1$. $\rho_{xy}$ and $\rho_{yx}$ are proportional to $B$ originating from the second term of Eqs.~(\ref{Lorenznumberxx},~\ref{Lorenznumberyy}). As to the Seebeck coefficient, however, the linear dependence of $S$ on $B$ is cancelled due to the definition.

To illustrate the efficiency of the magnetothermoelectric conversion, here we define the magnetoelectrical resistivity (MER), the magnetothermal resistivity (MTR) and the magnetoseebeck coefficient (MS) according to
\begin{subequations}
\begin{equation}
{\rm MER}=\frac{\rho(B\neq 0)-\rho(B=0)}{\rho(B=0)}
\end{equation}
\begin{equation}
{\rm MTR}=\frac{r(B\neq 0)-r(B=0)}{r(B=0)}
\end{equation}
\begin{equation}
{\rm MS}=\frac{S^{-1}(B\neq 0)-S^{-1}(B=0)}{S^{-1}(B=0)}
\end{equation}
\end{subequations}
where $S^{-1}$ is the inverse of the Seebeck coefficient. We can see from Fig.~4(d) that the magnetoelectrical resistivity increases exponentially with $B$. At $B=10$ Tesla, MER reaches up to 20\% approximately. When $B=50$ Tesla, ${\rm MER}\approx 40\%$ for $yy$ component. In Fig.~4(e) we show the magnetothermal resistivity as a function of magnetic field. It is found that at $B<10$ Tesla, MTR varies slowly. Nevertheless, once $B>50$ Tesla, MTR increases sharply and ${\rm MTR}\approx 500\%$ at $B=80$ Tesla. By reducing the temperature, our calculations reveal that MER and MTR can be further enhanced and the corresponding results are shown in Fig.~S7 [see SI]. We obtain a quite large MER and MTR even at small $B$. The origin is that at low temperatures less scattering occurs and magnetic field effect becomes obvious. In Fig.~4(f) we plot the magnetoseebeck coefficient versus the magnetic field. It is shown that MS increases slightly at the beginning and then decreases with $B$. Although the ratio of MS is not appreciable, the large MER and MTR suggest a great potential for device applications of SnSe in the field of magnetic memory.


In conclusion, we have employed the deformation potential theory to investigate the effect of EPC on the thermoelectric transport properties of 2D SnSe sheet. We found: (i) the EPC strength in this 2D material is especially large in contrast to that of the pristine graphene; (ii) the scattering rate depends intensively on the system temperatures and the carrier densities when the Fermi energy is close to the band edges, indicating the deficiency of the constant relaxation time approximation. The scattering rates as a function of temperature for both acoustic and optical phonon modes can be fitted excellently by a polynomial formula. The shear modes' contributions for transport compared to the longitudinal ones are negligible. Based on the Boltzmann transport equations combined with the first-principle calculations, we investigated the magnetothermoelectric effect of the 2D SnSe. It is shown that at low temperatures there are enormous magnetoelectrical resistivity and magnetothermal resistivity which can be further enhanced by reducing the temperature. Our results compare reasonably well with the experimental data. This study qualifies 2D SnSe as an outstanding EPC material and a compelling magnetothermoelectric material. We expect that these findings are beneficial for the future in the nano-device applications.

{\bf Supporting Information.} \\
The Supporting Information is available free of charge on the ACS Publications website at DOI: XXXXX.

The tilted view of the single-layer SnSe sheet; PDOS for each component of SnSe; A schematic illustration of the dilation of the lattice in calculation of the deformation potentials; Deformation potentials: $\Xi_y$ and ${\mathcal D}_y$ for the LA and LO phonon branches; Energy bands of the 2D SnSe with and without lattice distortions; A comparison of the transport coefficients between theory and experiment; The magnetothermoelectric coefficients at temperature $T=50$ K (PDF).

{\bf Corresponding Author.} \\
E-mail: phyyk@nus.edu.sg (K.Y.)

{\bf ORCID.} \\
Kaike Yang: 0000-0001-7482-4019

{\bf Notes.} \\
The authors declare no competing financial interest.

{\bf Acknowledgements.} \\
The authors acknowledge L.~D.~Zhao for providing the experimental data. This work was supported by the FRC grant (No.~R-144-000-343-112) and the MOE grant (No.~R-144-000-349-112) and the computational time was granted by the National Supercomputing Centre of Singapore.

\bibliography{SnSe.bib}

\begin{thebibliography}{56}%
\makeatletter
\providecommand \@ifxundefined [1]{%
 \@ifx{#1\undefined}
}%
\providecommand \@ifnum [1]{%
 \ifnum #1\expandafter \@firstoftwo
 \else \expandafter \@secondoftwo
 \fi
}%
\providecommand \@ifx [1]{%
 \ifx #1\expandafter \@firstoftwo
 \else \expandafter \@secondoftwo
 \fi
}%
\providecommand \natexlab [1]{#1}%
\providecommand \enquote  [1]{``#1''}%
\providecommand \bibnamefont  [1]{#1}%
\providecommand \bibfnamefont [1]{#1}%
\providecommand \citenamefont [1]{#1}%
\providecommand \href@noop [0]{\@secondoftwo}%
\providecommand \href [0]{\begingroup \@sanitize@url \@href}%
\providecommand \@href[1]{\@@startlink{#1}\@@href}%
\providecommand \@@href[1]{\endgroup#1\@@endlink}%
\providecommand \@sanitize@url [0]{\catcode `\\12\catcode `\$12\catcode
  `\&12\catcode `\#12\catcode `\^12\catcode `\_12\catcode `\%12\relax}%
\providecommand \@@startlink[1]{}%
\providecommand \@@endlink[0]{}%
\providecommand \url  [0]{\begingroup\@sanitize@url \@url }%
\providecommand \@url [1]{\endgroup\@href {#1}{\urlprefix }}%
\providecommand \urlprefix  [0]{URL }%
\providecommand \Eprint [0]{\href }%
\providecommand \doibase [0]{http://dx.doi.org/}%
\providecommand \selectlanguage [0]{\@gobble}%
\providecommand \bibinfo  [0]{\@secondoftwo}%
\providecommand \bibfield  [0]{\@secondoftwo}%
\providecommand \translation [1]{[#1]}%
\providecommand \BibitemOpen [0]{}%
\providecommand \bibitemStop [0]{}%
\providecommand \bibitemNoStop [0]{.\EOS\space}%
\providecommand \EOS [0]{\spacefactor3000\relax}%
\providecommand \BibitemShut  [1]{\csname bibitem#1\endcsname}%
\let\auto@bib@innerbib\@empty
\bibitem [{\citenamefont {Goldsmid}(2010)}]{Goldsmid2010}%
  \BibitemOpen
  \bibfield  {author} {\bibinfo {author} {\bibfnamefont {H.~J.}\ \bibnamefont
  {Goldsmid}},\ }\href {\doibase 10.1007/978-3-642-00716-3} {\emph {\bibinfo
  {title} {{Introduction to Thermoelectricity}}}}\ (\bibinfo  {publisher}
  {Springer Berlin Heidelberg},\ \bibinfo {address} {Berlin, Heidelberg},\
  \bibinfo {year} {2010})\BibitemShut {NoStop}%
\bibitem [{\citenamefont {DiSalvo}(1999)}]{Disalvo1999}%
  \BibitemOpen
  \bibfield  {author} {\bibinfo {author} {\bibfnamefont {F.~J.}\ \bibnamefont
  {DiSalvo}},\ }\href {\doibase 10.1126/science.285.5428.703} {\bibfield
  {journal} {\bibinfo  {journal} {Science}\ }\textbf {\bibinfo {volume}
  {285}},\ \bibinfo {pages} {703} (\bibinfo {year} {1999})}\BibitemShut
  {NoStop}%
\bibitem [{\citenamefont {Vining}(2008)}]{Vining2008}%
  \BibitemOpen
  \bibfield  {author} {\bibinfo {author} {\bibfnamefont {C.~B.}\ \bibnamefont
  {Vining}},\ }\href {\doibase 10.1038/nmat2271} {\bibfield  {journal}
  {\bibinfo  {journal} {Nat. Mater.}\ }\textbf {\bibinfo {volume} {7}},\
  \bibinfo {pages} {765} (\bibinfo {year} {2008})}\BibitemShut {NoStop}%
\bibitem [{\citenamefont {Dubi}\ and\ \citenamefont {{Di
  Ventra}}(2011)}]{Dubi2011}%
  \BibitemOpen
  \bibfield  {author} {\bibinfo {author} {\bibfnamefont {Y.}~\bibnamefont
  {Dubi}}\ and\ \bibinfo {author} {\bibfnamefont {M.}~\bibnamefont {{Di
  Ventra}}},\ }\href {\doibase 10.1103/RevModPhys.83.131} {\bibfield  {journal}
  {\bibinfo  {journal} {Rev. Mod. Phys.}\ }\textbf {\bibinfo {volume} {83}},\
  \bibinfo {pages} {131} (\bibinfo {year} {2011})}\BibitemShut {NoStop}%
\bibitem [{\citenamefont {Yang}\ \emph {et~al.}(2012)\citenamefont {Yang},
  \citenamefont {Chen}, \citenamefont {D'Agosta}, \citenamefont {Xie},
  \citenamefont {Zhong},\ and\ \citenamefont {Rubio}}]{Yang2012}%
  \BibitemOpen
  \bibfield  {author} {\bibinfo {author} {\bibfnamefont {K.}~\bibnamefont
  {Yang}}, \bibinfo {author} {\bibfnamefont {Y.}~\bibnamefont {Chen}}, \bibinfo
  {author} {\bibfnamefont {R.}~\bibnamefont {D'Agosta}}, \bibinfo {author}
  {\bibfnamefont {Y.}~\bibnamefont {Xie}}, \bibinfo {author} {\bibfnamefont
  {J.}~\bibnamefont {Zhong}}, \ and\ \bibinfo {author} {\bibfnamefont
  {A.}~\bibnamefont {Rubio}},\ }\href {\doibase 10.1103/PhysRevB.86.045425}
  {\bibfield  {journal} {\bibinfo  {journal} {Phys. Rev. B}\ }\textbf {\bibinfo
  {volume} {86}},\ \bibinfo {pages} {045425} (\bibinfo {year}
  {2012})}\BibitemShut {NoStop}%
\bibitem [{\citenamefont {D'Agosta}(2013)}]{Dagosta2013}%
  \BibitemOpen
  \bibfield  {author} {\bibinfo {author} {\bibfnamefont {R.}~\bibnamefont
  {D'Agosta}},\ }\href {\doibase 10.1039/C2CP42594G} {\bibfield  {journal}
  {\bibinfo  {journal} {Phys. Chem. Chem. Phys.}\ }\textbf {\bibinfo {volume}
  {15}},\ \bibinfo {pages} {1758} (\bibinfo {year} {2013})}\BibitemShut
  {NoStop}%
\bibitem [{\citenamefont {Heremans}\ \emph {et~al.}(2008)\citenamefont
  {Heremans}, \citenamefont {Jovovic}, \citenamefont {Toberer}, \citenamefont
  {Saramat}, \citenamefont {Kurosaki}, \citenamefont {Charoenphakdee},
  \citenamefont {Yamanaka},\ and\ \citenamefont {Snyder}}]{Heremans2008}%
  \BibitemOpen
  \bibfield  {author} {\bibinfo {author} {\bibfnamefont {J.~P.}\ \bibnamefont
  {Heremans}}, \bibinfo {author} {\bibfnamefont {V.}~\bibnamefont {Jovovic}},
  \bibinfo {author} {\bibfnamefont {E.~S.}\ \bibnamefont {Toberer}}, \bibinfo
  {author} {\bibfnamefont {A.}~\bibnamefont {Saramat}}, \bibinfo {author}
  {\bibfnamefont {K.}~\bibnamefont {Kurosaki}}, \bibinfo {author}
  {\bibfnamefont {A.}~\bibnamefont {Charoenphakdee}}, \bibinfo {author}
  {\bibfnamefont {S.}~\bibnamefont {Yamanaka}}, \ and\ \bibinfo {author}
  {\bibfnamefont {G.~J.}\ \bibnamefont {Snyder}},\ }\href {\doibase
  10.1126/science.1159725} {\bibfield  {journal} {\bibinfo  {journal}
  {Science}\ }\textbf {\bibinfo {volume} {321}},\ \bibinfo {pages} {554}
  (\bibinfo {year} {2008})}\BibitemShut {NoStop}%
\bibitem [{\citenamefont {Hicks}\ and\ \citenamefont
  {Dresselhaus}(1993{\natexlab{a}})}]{Hicks1993}%
  \BibitemOpen
  \bibfield  {author} {\bibinfo {author} {\bibfnamefont {L.~D.}\ \bibnamefont
  {Hicks}}\ and\ \bibinfo {author} {\bibfnamefont {M.~S.}\ \bibnamefont
  {Dresselhaus}},\ }\href {\doibase 10.1103/PhysRevB.47.12727} {\bibfield
  {journal} {\bibinfo  {journal} {Phys. Rev. B}\ }\textbf {\bibinfo {volume}
  {47}},\ \bibinfo {pages} {12727} (\bibinfo {year}
  {1993}{\natexlab{a}})}\BibitemShut {NoStop}%
\bibitem [{\citenamefont {Hicks}\ and\ \citenamefont
  {Dresselhaus}(1993{\natexlab{b}})}]{Hicks1993a}%
  \BibitemOpen
  \bibfield  {author} {\bibinfo {author} {\bibfnamefont {L.~D.}\ \bibnamefont
  {Hicks}}\ and\ \bibinfo {author} {\bibfnamefont {M.~S.}\ \bibnamefont
  {Dresselhaus}},\ }\href {\doibase 10.1103/PhysRevB.47.16631} {\bibfield
  {journal} {\bibinfo  {journal} {Phys. Rev. B}\ }\textbf {\bibinfo {volume}
  {47}},\ \bibinfo {pages} {16631} (\bibinfo {year}
  {1993}{\natexlab{b}})}\BibitemShut {NoStop}%
\bibitem [{\citenamefont {Mahan}\ and\ \citenamefont {Sofo}(1996)}]{Mahan1996}%
  \BibitemOpen
  \bibfield  {author} {\bibinfo {author} {\bibfnamefont {G.~D.}\ \bibnamefont
  {Mahan}}\ and\ \bibinfo {author} {\bibfnamefont {J.~O.}\ \bibnamefont
  {Sofo}},\ }\href {\doibase 10.1073/pnas.93.15.7436} {\bibfield  {journal}
  {\bibinfo  {journal} {Proc. Natl. Acad. Sci.}\ }\textbf {\bibinfo {volume}
  {93}},\ \bibinfo {pages} {7436} (\bibinfo {year} {1996})}\BibitemShut
  {NoStop}%
\bibitem [{\citenamefont {Hochbaum}\ \emph {et~al.}(2008)\citenamefont
  {Hochbaum}, \citenamefont {Chen}, \citenamefont {Delgado}, \citenamefont
  {Liang}, \citenamefont {Garnett}, \citenamefont {Najarian}, \citenamefont
  {Majumdar},\ and\ \citenamefont {Yang}}]{Hochbaum2008}%
  \BibitemOpen
  \bibfield  {author} {\bibinfo {author} {\bibfnamefont {A.~I.}\ \bibnamefont
  {Hochbaum}}, \bibinfo {author} {\bibfnamefont {R.}~\bibnamefont {Chen}},
  \bibinfo {author} {\bibfnamefont {R.~D.}\ \bibnamefont {Delgado}}, \bibinfo
  {author} {\bibfnamefont {W.}~\bibnamefont {Liang}}, \bibinfo {author}
  {\bibfnamefont {E.~C.}\ \bibnamefont {Garnett}}, \bibinfo {author}
  {\bibfnamefont {M.}~\bibnamefont {Najarian}}, \bibinfo {author}
  {\bibfnamefont {A.}~\bibnamefont {Majumdar}}, \ and\ \bibinfo {author}
  {\bibfnamefont {P.}~\bibnamefont {Yang}},\ }\href {\doibase
  10.1038/nature06381} {\bibfield  {journal} {\bibinfo  {journal} {Nature}\
  }\textbf {\bibinfo {volume} {451}},\ \bibinfo {pages} {163} (\bibinfo {year}
  {2008})}\BibitemShut {NoStop}%
\bibitem [{\citenamefont {Boukai}\ \emph {et~al.}(2008)\citenamefont {Boukai},
  \citenamefont {Bunimovich}, \citenamefont {Tahir-Kheli}, \citenamefont {Yu},
  \citenamefont {{Goddard III}},\ and\ \citenamefont {Heath}}]{Boukai2008}%
  \BibitemOpen
  \bibfield  {author} {\bibinfo {author} {\bibfnamefont {A.~I.}\ \bibnamefont
  {Boukai}}, \bibinfo {author} {\bibfnamefont {Y.}~\bibnamefont {Bunimovich}},
  \bibinfo {author} {\bibfnamefont {J.}~\bibnamefont {Tahir-Kheli}}, \bibinfo
  {author} {\bibfnamefont {J.~K.}\ \bibnamefont {Yu}}, \bibinfo {author}
  {\bibfnamefont {W.~A.}\ \bibnamefont {{Goddard III}}}, \ and\ \bibinfo
  {author} {\bibfnamefont {J.~R.}\ \bibnamefont {Heath}},\ }\href {\doibase
  10.1038/nature06458} {\bibfield  {journal} {\bibinfo  {journal} {Nature}\
  }\textbf {\bibinfo {volume} {451}},\ \bibinfo {pages} {168} (\bibinfo {year}
  {2008})}\BibitemShut {NoStop}%
\bibitem [{\citenamefont {Ni}\ \emph {et~al.}(2009)\citenamefont {Ni},
  \citenamefont {Liang}, \citenamefont {Wang},\ and\ \citenamefont
  {Li}}]{Ni2009}%
  \BibitemOpen
  \bibfield  {author} {\bibinfo {author} {\bibfnamefont {X.}~\bibnamefont
  {Ni}}, \bibinfo {author} {\bibfnamefont {G.}~\bibnamefont {Liang}}, \bibinfo
  {author} {\bibfnamefont {J.-S.}\ \bibnamefont {Wang}}, \ and\ \bibinfo
  {author} {\bibfnamefont {B.}~\bibnamefont {Li}},\ }\href {\doibase
  10.1063/1.3264087} {\bibfield  {journal} {\bibinfo  {journal} {Appl. Phys.
  Lett.}\ }\textbf {\bibinfo {volume} {95}},\ \bibinfo {pages} {192114}
  (\bibinfo {year} {2009})}\BibitemShut {NoStop}%
\bibitem [{\citenamefont {Yang}\ \emph {et~al.}(2014)\citenamefont {Yang},
  \citenamefont {Cahangirov}, \citenamefont {Cantarero}, \citenamefont
  {Rubio},\ and\ \citenamefont {D'Agosta}}]{Yang2014}%
  \BibitemOpen
  \bibfield  {author} {\bibinfo {author} {\bibfnamefont {K.}~\bibnamefont
  {Yang}}, \bibinfo {author} {\bibfnamefont {S.}~\bibnamefont {Cahangirov}},
  \bibinfo {author} {\bibfnamefont {A.}~\bibnamefont {Cantarero}}, \bibinfo
  {author} {\bibfnamefont {A.}~\bibnamefont {Rubio}}, \ and\ \bibinfo {author}
  {\bibfnamefont {R.}~\bibnamefont {D'Agosta}},\ }\href {\doibase
  10.1103/PhysRevB.89.125403} {\bibfield  {journal} {\bibinfo  {journal} {Phys.
  Rev. B}\ }\textbf {\bibinfo {volume} {89}},\ \bibinfo {pages} {125403}
  (\bibinfo {year} {2014})}\BibitemShut {NoStop}%
\bibitem [{\citenamefont {Yang}\ \emph {et~al.}(2015)\citenamefont {Yang},
  \citenamefont {Cantarero}, \citenamefont {Rubio},\ and\ \citenamefont
  {D'Agosta}}]{Yang2015}%
  \BibitemOpen
  \bibfield  {author} {\bibinfo {author} {\bibfnamefont {K.}~\bibnamefont
  {Yang}}, \bibinfo {author} {\bibfnamefont {A.}~\bibnamefont {Cantarero}},
  \bibinfo {author} {\bibfnamefont {A.}~\bibnamefont {Rubio}}, \ and\ \bibinfo
  {author} {\bibfnamefont {R.}~\bibnamefont {D'Agosta}},\ }\href {\doibase
  10.1007/s12274-015-0766-2} {\bibfield  {journal} {\bibinfo  {journal} {Nano
  Res.}\ }\textbf {\bibinfo {volume} {8}},\ \bibinfo {pages} {2611} (\bibinfo
  {year} {2015})}\BibitemShut {NoStop}%
\bibitem [{\citenamefont {Chattopadhyay}\ \emph {et~al.}(1986)\citenamefont
  {Chattopadhyay}, \citenamefont {Pannetier},\ and\ \citenamefont {{Von
  Schnering}}}]{Chattopadhyay1986a}%
  \BibitemOpen
  \bibfield  {author} {\bibinfo {author} {\bibfnamefont {T.}~\bibnamefont
  {Chattopadhyay}}, \bibinfo {author} {\bibfnamefont {J.}~\bibnamefont
  {Pannetier}}, \ and\ \bibinfo {author} {\bibfnamefont {H.}~\bibnamefont {{Von
  Schnering}}},\ }\href {\doibase 10.1016/0022-3697(86)90059-4} {\bibfield
  {journal} {\bibinfo  {journal} {J. Phys. Chem. Solids}\ }\textbf {\bibinfo
  {volume} {47}},\ \bibinfo {pages} {879} (\bibinfo {year} {1986})}\BibitemShut
  {NoStop}%
\bibitem [{\citenamefont {Zhao}\ \emph {et~al.}(2014)\citenamefont {Zhao},
  \citenamefont {Lo}, \citenamefont {Zhang}, \citenamefont {Sun}, \citenamefont
  {Tan}, \citenamefont {Uher}, \citenamefont {Wolverton}, \citenamefont
  {Dravid},\ and\ \citenamefont {Kanatzidis}}]{Zhao2014}%
  \BibitemOpen
  \bibfield  {author} {\bibinfo {author} {\bibfnamefont {L.~D.}\ \bibnamefont
  {Zhao}}, \bibinfo {author} {\bibfnamefont {S.~H.}\ \bibnamefont {Lo}},
  \bibinfo {author} {\bibfnamefont {Y.}~\bibnamefont {Zhang}}, \bibinfo
  {author} {\bibfnamefont {H.}~\bibnamefont {Sun}}, \bibinfo {author}
  {\bibfnamefont {G.}~\bibnamefont {Tan}}, \bibinfo {author} {\bibfnamefont
  {C.}~\bibnamefont {Uher}}, \bibinfo {author} {\bibfnamefont {C.}~\bibnamefont
  {Wolverton}}, \bibinfo {author} {\bibfnamefont {V.~P.}\ \bibnamefont
  {Dravid}}, \ and\ \bibinfo {author} {\bibfnamefont {M.~G.}\ \bibnamefont
  {Kanatzidis}},\ }\href {\doibase 10.1038/nature13184} {\bibfield  {journal}
  {\bibinfo  {journal} {Nature}\ }\textbf {\bibinfo {volume} {508}},\ \bibinfo
  {pages} {373} (\bibinfo {year} {2014})}\BibitemShut {NoStop}%
\bibitem [{\citenamefont {Zhao}\ \emph
  {et~al.}(2016{\natexlab{a}})\citenamefont {Zhao}, \citenamefont {Tan},
  \citenamefont {Hao}, \citenamefont {He}, \citenamefont {Pei}, \citenamefont
  {Chi}, \citenamefont {Wang}, \citenamefont {Gong}, \citenamefont {Xu},
  \citenamefont {Dravid}, \citenamefont {Uher}, \citenamefont {Snyder},
  \citenamefont {Wolverton},\ and\ \citenamefont {Kanatzidis}}]{Zhao2016a}%
  \BibitemOpen
  \bibfield  {author} {\bibinfo {author} {\bibfnamefont {L.~D.}\ \bibnamefont
  {Zhao}}, \bibinfo {author} {\bibfnamefont {G.}~\bibnamefont {Tan}}, \bibinfo
  {author} {\bibfnamefont {S.}~\bibnamefont {Hao}}, \bibinfo {author}
  {\bibfnamefont {J.}~\bibnamefont {He}}, \bibinfo {author} {\bibfnamefont
  {Y.}~\bibnamefont {Pei}}, \bibinfo {author} {\bibfnamefont {H.}~\bibnamefont
  {Chi}}, \bibinfo {author} {\bibfnamefont {H.}~\bibnamefont {Wang}}, \bibinfo
  {author} {\bibfnamefont {S.}~\bibnamefont {Gong}}, \bibinfo {author}
  {\bibfnamefont {H.}~\bibnamefont {Xu}}, \bibinfo {author} {\bibfnamefont
  {V.~P.}\ \bibnamefont {Dravid}}, \bibinfo {author} {\bibfnamefont
  {C.}~\bibnamefont {Uher}}, \bibinfo {author} {\bibfnamefont {G.~J.}\
  \bibnamefont {Snyder}}, \bibinfo {author} {\bibfnamefont {C.}~\bibnamefont
  {Wolverton}}, \ and\ \bibinfo {author} {\bibfnamefont {M.~G.}\ \bibnamefont
  {Kanatzidis}},\ }\href {\doibase 10.1126/science.aad3749} {\bibfield
  {journal} {\bibinfo  {journal} {Science}\ }\textbf {\bibinfo {volume}
  {351}},\ \bibinfo {pages} {141} (\bibinfo {year}
  {2016}{\natexlab{a}})}\BibitemShut {NoStop}%
\bibitem [{\citenamefont {Ibrahim}\ \emph {et~al.}(2017)\citenamefont
  {Ibrahim}, \citenamefont {Vaney}, \citenamefont {Sassi}, \citenamefont
  {Candolfi}, \citenamefont {Ohorodniichuk}, \citenamefont {Levinsky},
  \citenamefont {Semprimoschnig}, \citenamefont {Dauscher},\ and\ \citenamefont
  {Lenoir}}]{Ibrahim2017}%
  \BibitemOpen
  \bibfield  {author} {\bibinfo {author} {\bibfnamefont {D.}~\bibnamefont
  {Ibrahim}}, \bibinfo {author} {\bibfnamefont {J.~B.}\ \bibnamefont {Vaney}},
  \bibinfo {author} {\bibfnamefont {S.}~\bibnamefont {Sassi}}, \bibinfo
  {author} {\bibfnamefont {C.}~\bibnamefont {Candolfi}}, \bibinfo {author}
  {\bibfnamefont {V.}~\bibnamefont {Ohorodniichuk}}, \bibinfo {author}
  {\bibfnamefont {P.}~\bibnamefont {Levinsky}}, \bibinfo {author}
  {\bibfnamefont {C.}~\bibnamefont {Semprimoschnig}}, \bibinfo {author}
  {\bibfnamefont {A.}~\bibnamefont {Dauscher}}, \ and\ \bibinfo {author}
  {\bibfnamefont {B.}~\bibnamefont {Lenoir}},\ }\href {\doibase
  10.1063/1.4974348} {\bibfield  {journal} {\bibinfo  {journal} {Appl. Phys.
  Lett.}\ }\textbf {\bibinfo {volume} {110}},\ \bibinfo {pages} {032103}
  (\bibinfo {year} {2017})}\BibitemShut {NoStop}%
\bibitem [{\citenamefont {Carrete}\ \emph {et~al.}(2014)\citenamefont
  {Carrete}, \citenamefont {Mingo},\ and\ \citenamefont
  {Curtarolo}}]{Carrete2014}%
  \BibitemOpen
  \bibfield  {author} {\bibinfo {author} {\bibfnamefont {J.}~\bibnamefont
  {Carrete}}, \bibinfo {author} {\bibfnamefont {N.}~\bibnamefont {Mingo}}, \
  and\ \bibinfo {author} {\bibfnamefont {S.}~\bibnamefont {Curtarolo}},\ }\href
  {\doibase 10.1063/1.4895770} {\bibfield  {journal} {\bibinfo  {journal}
  {Appl. Phys. Lett.}\ }\textbf {\bibinfo {volume} {105}},\ \bibinfo {pages}
  {101907} (\bibinfo {year} {2014})}\BibitemShut {NoStop}%
\bibitem [{\citenamefont {Skelton}\ \emph {et~al.}(2016)\citenamefont
  {Skelton}, \citenamefont {Burton}, \citenamefont {Parker}, \citenamefont
  {Walsh}, \citenamefont {Kim}, \citenamefont {Soon}, \citenamefont
  {Buckeridge}, \citenamefont {Sokol}, \citenamefont {Catlow}, \citenamefont
  {Togo},\ and\ \citenamefont {Tanaka}}]{Skelton2016}%
  \BibitemOpen
  \bibfield  {author} {\bibinfo {author} {\bibfnamefont {J.~M.}\ \bibnamefont
  {Skelton}}, \bibinfo {author} {\bibfnamefont {L.~A.}\ \bibnamefont {Burton}},
  \bibinfo {author} {\bibfnamefont {S.~C.}\ \bibnamefont {Parker}}, \bibinfo
  {author} {\bibfnamefont {A.}~\bibnamefont {Walsh}}, \bibinfo {author}
  {\bibfnamefont {C.-E.}\ \bibnamefont {Kim}}, \bibinfo {author} {\bibfnamefont
  {A.}~\bibnamefont {Soon}}, \bibinfo {author} {\bibfnamefont {J.}~\bibnamefont
  {Buckeridge}}, \bibinfo {author} {\bibfnamefont {A.~A.}\ \bibnamefont
  {Sokol}}, \bibinfo {author} {\bibfnamefont {C.~R.~A.}\ \bibnamefont
  {Catlow}}, \bibinfo {author} {\bibfnamefont {A.}~\bibnamefont {Togo}}, \ and\
  \bibinfo {author} {\bibfnamefont {I.}~\bibnamefont {Tanaka}},\ }\href
  {\doibase 10.1103/PhysRevLett.117.075502} {\bibfield  {journal} {\bibinfo
  {journal} {Phys. Rev. Lett.}\ }\textbf {\bibinfo {volume} {117}},\ \bibinfo
  {pages} {075502} (\bibinfo {year} {2016})}\BibitemShut {NoStop}%
\bibitem [{\citenamefont {Li}\ \emph {et~al.}(2015)\citenamefont {Li},
  \citenamefont {Hong}, \citenamefont {May}, \citenamefont {Bansal},
  \citenamefont {Chi}, \citenamefont {Hong}, \citenamefont {Ehlers},\ and\
  \citenamefont {Delaire}}]{Li2015}%
  \BibitemOpen
  \bibfield  {author} {\bibinfo {author} {\bibfnamefont {C.~W.}\ \bibnamefont
  {Li}}, \bibinfo {author} {\bibfnamefont {J.}~\bibnamefont {Hong}}, \bibinfo
  {author} {\bibfnamefont {A.~F.}\ \bibnamefont {May}}, \bibinfo {author}
  {\bibfnamefont {D.}~\bibnamefont {Bansal}}, \bibinfo {author} {\bibfnamefont
  {S.}~\bibnamefont {Chi}}, \bibinfo {author} {\bibfnamefont {T.}~\bibnamefont
  {Hong}}, \bibinfo {author} {\bibfnamefont {G.}~\bibnamefont {Ehlers}}, \ and\
  \bibinfo {author} {\bibfnamefont {O.}~\bibnamefont {Delaire}},\ }\href
  {\doibase 10.1038/nphys3492} {\bibfield  {journal} {\bibinfo  {journal} {Nat.
  Phys.}\ }\textbf {\bibinfo {volume} {11}},\ \bibinfo {pages} {1063} (\bibinfo
  {year} {2015})}\BibitemShut {NoStop}%
\bibitem [{\citenamefont {Bardeen}\ and\ \citenamefont
  {Shockley}(1950)}]{Bardeen1950}%
  \BibitemOpen
  \bibfield  {author} {\bibinfo {author} {\bibfnamefont {J.}~\bibnamefont
  {Bardeen}}\ and\ \bibinfo {author} {\bibfnamefont {W.}~\bibnamefont
  {Shockley}},\ }\href {\doibase 10.1103/PhysRev.80.72} {\bibfield  {journal}
  {\bibinfo  {journal} {Phys. Rev.}\ }\textbf {\bibinfo {volume} {80}},\
  \bibinfo {pages} {72} (\bibinfo {year} {1950})}\BibitemShut {NoStop}%
\bibitem [{\citenamefont {Herring}\ and\ \citenamefont
  {Vogt}(1956)}]{Herring1956}%
  \BibitemOpen
  \bibfield  {author} {\bibinfo {author} {\bibfnamefont {C.}~\bibnamefont
  {Herring}}\ and\ \bibinfo {author} {\bibfnamefont {E.}~\bibnamefont {Vogt}},\
  }\href {\doibase 10.1103/PhysRev.101.944} {\bibfield  {journal} {\bibinfo
  {journal} {Phys. Rev.}\ }\textbf {\bibinfo {volume} {101}},\ \bibinfo {pages}
  {944} (\bibinfo {year} {1956})}\BibitemShut {NoStop}%
\bibitem [{\citenamefont {Gantmakher}\ and\ \citenamefont
  {Levinson}(1987)}]{Gantmakher1987}%
  \BibitemOpen
  \bibfield  {author} {\bibinfo {author} {\bibfnamefont {V.~F.}\ \bibnamefont
  {Gantmakher}}\ and\ \bibinfo {author} {\bibfnamefont {Y.~B.}\ \bibnamefont
  {Levinson}},\ }\href {\doibase 10.1063/1.2811285} {\emph {\bibinfo {title}
  {{Carrier Scattering in Metals and Semiconductors}}}}\ (\bibinfo  {publisher}
  {North-Holland Physics Publishing},\ \bibinfo {address} {The Netherlands},\
  \bibinfo {year} {1987})\BibitemShut {NoStop}%
\bibitem [{\citenamefont {Ulstrup}\ \emph {et~al.}(2012)\citenamefont
  {Ulstrup}, \citenamefont {Bianchi}, \citenamefont {Hatch}, \citenamefont
  {Guan}, \citenamefont {Baraldi}, \citenamefont {Alf{\`{e}}}, \citenamefont
  {Hornek{\ae}r},\ and\ \citenamefont {Hofmann}}]{Ulstrup2012}%
  \BibitemOpen
  \bibfield  {author} {\bibinfo {author} {\bibfnamefont {S.}~\bibnamefont
  {Ulstrup}}, \bibinfo {author} {\bibfnamefont {M.}~\bibnamefont {Bianchi}},
  \bibinfo {author} {\bibfnamefont {R.}~\bibnamefont {Hatch}}, \bibinfo
  {author} {\bibfnamefont {D.}~\bibnamefont {Guan}}, \bibinfo {author}
  {\bibfnamefont {A.}~\bibnamefont {Baraldi}}, \bibinfo {author} {\bibfnamefont
  {D.}~\bibnamefont {Alf{\`{e}}}}, \bibinfo {author} {\bibfnamefont
  {L.}~\bibnamefont {Hornek{\ae}r}}, \ and\ \bibinfo {author} {\bibfnamefont
  {P.}~\bibnamefont {Hofmann}},\ }\href {\doibase 10.1103/PhysRevB.86.161402}
  {\bibfield  {journal} {\bibinfo  {journal} {Phys. Rev. B}\ }\textbf {\bibinfo
  {volume} {86}},\ \bibinfo {pages} {161402} (\bibinfo {year}
  {2012})}\BibitemShut {NoStop}%
\bibitem [{\citenamefont {Dewandre}\ \emph {et~al.}(2016)\citenamefont
  {Dewandre}, \citenamefont {Hellman}, \citenamefont {Bhattacharya},
  \citenamefont {Romero}, \citenamefont {Madsen},\ and\ \citenamefont
  {Verstraete}}]{Dewandre2016a}%
  \BibitemOpen
  \bibfield  {author} {\bibinfo {author} {\bibfnamefont {A.}~\bibnamefont
  {Dewandre}}, \bibinfo {author} {\bibfnamefont {O.}~\bibnamefont {Hellman}},
  \bibinfo {author} {\bibfnamefont {S.}~\bibnamefont {Bhattacharya}}, \bibinfo
  {author} {\bibfnamefont {A.~H.}\ \bibnamefont {Romero}}, \bibinfo {author}
  {\bibfnamefont {G.~K.~H.}\ \bibnamefont {Madsen}}, \ and\ \bibinfo {author}
  {\bibfnamefont {M.~J.}\ \bibnamefont {Verstraete}},\ }\href {\doibase
  10.1103/PhysRevLett.117.276601} {\bibfield  {journal} {\bibinfo  {journal}
  {Phys. Rev. Lett.}\ }\textbf {\bibinfo {volume} {117}},\ \bibinfo {pages}
  {276601} (\bibinfo {year} {2016})}\BibitemShut {NoStop}%
\bibitem [{\citenamefont {Bansal}\ \emph {et~al.}(2016)\citenamefont {Bansal},
  \citenamefont {Hong}, \citenamefont {Li}, \citenamefont {May}, \citenamefont
  {Porter}, \citenamefont {Hu}, \citenamefont {Abernathy},\ and\ \citenamefont
  {Delaire}}]{Bansal2016a}%
  \BibitemOpen
  \bibfield  {author} {\bibinfo {author} {\bibfnamefont {D.}~\bibnamefont
  {Bansal}}, \bibinfo {author} {\bibfnamefont {J.}~\bibnamefont {Hong}},
  \bibinfo {author} {\bibfnamefont {C.~W.}\ \bibnamefont {Li}}, \bibinfo
  {author} {\bibfnamefont {A.~F.}\ \bibnamefont {May}}, \bibinfo {author}
  {\bibfnamefont {W.}~\bibnamefont {Porter}}, \bibinfo {author} {\bibfnamefont
  {M.~Y.}\ \bibnamefont {Hu}}, \bibinfo {author} {\bibfnamefont {D.~L.}\
  \bibnamefont {Abernathy}}, \ and\ \bibinfo {author} {\bibfnamefont
  {O.}~\bibnamefont {Delaire}},\ }\href {\doibase 10.1103/PhysRevB.94.054307}
  {\bibfield  {journal} {\bibinfo  {journal} {Phys. Rev. B}\ }\textbf {\bibinfo
  {volume} {94}},\ \bibinfo {pages} {054307} (\bibinfo {year}
  {2016})}\BibitemShut {NoStop}%
\bibitem [{\citenamefont {Hong}\ and\ \citenamefont
  {Delaire}(2016)}]{Hong2016a}%
  \BibitemOpen
  \bibfield  {author} {\bibinfo {author} {\bibfnamefont {J.}~\bibnamefont
  {Hong}}\ and\ \bibinfo {author} {\bibfnamefont {O.}~\bibnamefont {Delaire}},\
  }\href {http://arxiv.org/abs/1604.07077} {\  (\bibinfo {year} {2016})},\
  \Eprint {http://arxiv.org/abs/1604.07077} {arXiv:1604.07077} \BibitemShut
  {NoStop}%
\bibitem [{\citenamefont {Li}\ \emph {et~al.}(2013)\citenamefont {Li},
  \citenamefont {Chen}, \citenamefont {Hu}, \citenamefont {Wang}, \citenamefont
  {Zhang}, \citenamefont {Chen},\ and\ \citenamefont {Wang}}]{Li2013a}%
  \BibitemOpen
  \bibfield  {author} {\bibinfo {author} {\bibfnamefont {L.}~\bibnamefont
  {Li}}, \bibinfo {author} {\bibfnamefont {Z.}~\bibnamefont {Chen}}, \bibinfo
  {author} {\bibfnamefont {Y.}~\bibnamefont {Hu}}, \bibinfo {author}
  {\bibfnamefont {X.}~\bibnamefont {Wang}}, \bibinfo {author} {\bibfnamefont
  {T.}~\bibnamefont {Zhang}}, \bibinfo {author} {\bibfnamefont
  {W.}~\bibnamefont {Chen}}, \ and\ \bibinfo {author} {\bibfnamefont
  {Q.}~\bibnamefont {Wang}},\ }\href {\doibase 10.1021/ja3108017} {\bibfield
  {journal} {\bibinfo  {journal} {J. Am. Chem. Soc.}\ }\textbf {\bibinfo
  {volume} {135}},\ \bibinfo {pages} {1213} (\bibinfo {year}
  {2013})}\BibitemShut {NoStop}%
\bibitem [{\citenamefont {Kohn}\ and\ \citenamefont {Sham}(1965)}]{Kohn1965a}%
  \BibitemOpen
  \bibfield  {author} {\bibinfo {author} {\bibfnamefont {W.}~\bibnamefont
  {Kohn}}\ and\ \bibinfo {author} {\bibfnamefont {L.~J.}\ \bibnamefont
  {Sham}},\ }\href {\doibase 10.1103/PhysRev.140.A1133} {\bibfield  {journal}
  {\bibinfo  {journal} {Phys. Rev.}\ }\textbf {\bibinfo {volume} {140}},\
  \bibinfo {pages} {A1133} (\bibinfo {year} {1965})}\BibitemShut {NoStop}%
\bibitem [{\citenamefont {Giannozzi}\ \emph {et~al.}(2009)\citenamefont
  {Giannozzi}, \citenamefont {Baroni}, \citenamefont {Bonini}, \citenamefont
  {Calandra}, \citenamefont {Car}, \citenamefont {Cavazzoni}, \citenamefont
  {Ceresoli}, \citenamefont {Chiarotti}, \citenamefont {Cococcioni},
  \citenamefont {Dabo}, \citenamefont {{Dal Corso}}, \citenamefont
  {de~Gironcoli}, \citenamefont {Fabris}, \citenamefont {Fratesi},
  \citenamefont {Gebauer}, \citenamefont {Gerstmann}, \citenamefont
  {Gougoussis}, \citenamefont {Kokalj}, \citenamefont {Lazzeri}, \citenamefont
  {Martin-Samos}, \citenamefont {Marzari}, \citenamefont {Mauri}, \citenamefont
  {Mazzarello}, \citenamefont {Paolini}, \citenamefont {Pasquarello},
  \citenamefont {Paulatto}, \citenamefont {Sbraccia}, \citenamefont {Scandolo},
  \citenamefont {Sclauzero}, \citenamefont {Seitsonen}, \citenamefont
  {Smogunov}, \citenamefont {Umari},\ and\ \citenamefont
  {Wentzcovitch}}]{Giannozzi2009}%
  \BibitemOpen
  \bibfield  {author} {\bibinfo {author} {\bibfnamefont {P.}~\bibnamefont
  {Giannozzi}}, \bibinfo {author} {\bibfnamefont {S.}~\bibnamefont {Baroni}},
  \bibinfo {author} {\bibfnamefont {N.}~\bibnamefont {Bonini}}, \bibinfo
  {author} {\bibfnamefont {M.}~\bibnamefont {Calandra}}, \bibinfo {author}
  {\bibfnamefont {R.}~\bibnamefont {Car}}, \bibinfo {author} {\bibfnamefont
  {C.}~\bibnamefont {Cavazzoni}}, \bibinfo {author} {\bibfnamefont
  {D.}~\bibnamefont {Ceresoli}}, \bibinfo {author} {\bibfnamefont {G.~L.}\
  \bibnamefont {Chiarotti}}, \bibinfo {author} {\bibfnamefont {M.}~\bibnamefont
  {Cococcioni}}, \bibinfo {author} {\bibfnamefont {I.}~\bibnamefont {Dabo}},
  \bibinfo {author} {\bibfnamefont {A.}~\bibnamefont {{Dal Corso}}}, \bibinfo
  {author} {\bibfnamefont {S.}~\bibnamefont {de~Gironcoli}}, \bibinfo {author}
  {\bibfnamefont {S.}~\bibnamefont {Fabris}}, \bibinfo {author} {\bibfnamefont
  {G.}~\bibnamefont {Fratesi}}, \bibinfo {author} {\bibfnamefont
  {R.}~\bibnamefont {Gebauer}}, \bibinfo {author} {\bibfnamefont
  {U.}~\bibnamefont {Gerstmann}}, \bibinfo {author} {\bibfnamefont
  {C.}~\bibnamefont {Gougoussis}}, \bibinfo {author} {\bibfnamefont
  {A.}~\bibnamefont {Kokalj}}, \bibinfo {author} {\bibfnamefont
  {M.}~\bibnamefont {Lazzeri}}, \bibinfo {author} {\bibfnamefont
  {L.}~\bibnamefont {Martin-Samos}}, \bibinfo {author} {\bibfnamefont
  {N.}~\bibnamefont {Marzari}}, \bibinfo {author} {\bibfnamefont
  {F.}~\bibnamefont {Mauri}}, \bibinfo {author} {\bibfnamefont
  {R.}~\bibnamefont {Mazzarello}}, \bibinfo {author} {\bibfnamefont
  {S.}~\bibnamefont {Paolini}}, \bibinfo {author} {\bibfnamefont
  {A.}~\bibnamefont {Pasquarello}}, \bibinfo {author} {\bibfnamefont
  {L.}~\bibnamefont {Paulatto}}, \bibinfo {author} {\bibfnamefont
  {C.}~\bibnamefont {Sbraccia}}, \bibinfo {author} {\bibfnamefont
  {S.}~\bibnamefont {Scandolo}}, \bibinfo {author} {\bibfnamefont
  {G.}~\bibnamefont {Sclauzero}}, \bibinfo {author} {\bibfnamefont {A.~P.}\
  \bibnamefont {Seitsonen}}, \bibinfo {author} {\bibfnamefont {A.}~\bibnamefont
  {Smogunov}}, \bibinfo {author} {\bibfnamefont {P.}~\bibnamefont {Umari}}, \
  and\ \bibinfo {author} {\bibfnamefont {R.~M.}\ \bibnamefont {Wentzcovitch}},\
  }\href {\doibase 10.1088/0953-8984/21/39/395502} {\bibfield  {journal}
  {\bibinfo  {journal} {J. Phys. Condens. Matter}\ }\textbf {\bibinfo {volume}
  {21}},\ \bibinfo {pages} {395502} (\bibinfo {year} {2009})}\BibitemShut
  {NoStop}%
\bibitem [{\citenamefont {Perdew}\ \emph {et~al.}(1996)\citenamefont {Perdew},
  \citenamefont {Burke},\ and\ \citenamefont {Ernzerhof}}]{Perdew1996a}%
  \BibitemOpen
  \bibfield  {author} {\bibinfo {author} {\bibfnamefont {J.~P.}\ \bibnamefont
  {Perdew}}, \bibinfo {author} {\bibfnamefont {K.}~\bibnamefont {Burke}}, \
  and\ \bibinfo {author} {\bibfnamefont {M.}~\bibnamefont {Ernzerhof}},\ }\href
  {\doibase 10.1103/PhysRevLett.77.3865} {\bibfield  {journal} {\bibinfo
  {journal} {Phys. Rev. Lett.}\ }\textbf {\bibinfo {volume} {77}},\ \bibinfo
  {pages} {3865} (\bibinfo {year} {1996})}\BibitemShut {NoStop}%
\bibitem [{\citenamefont {Bl{\"{o}}chl}(1994)}]{Blochl1994}%
  \BibitemOpen
  \bibfield  {author} {\bibinfo {author} {\bibfnamefont {P.~E.}\ \bibnamefont
  {Bl{\"{o}}chl}},\ }\href {\doibase 10.1103/PhysRevB.50.17953} {\bibfield
  {journal} {\bibinfo  {journal} {Phys. Rev. B}\ }\textbf {\bibinfo {volume}
  {50}},\ \bibinfo {pages} {17953} (\bibinfo {year} {1994})}\BibitemShut
  {NoStop}%
\bibitem [{\citenamefont {Monkhorst}\ and\ \citenamefont
  {Pack}(1976)}]{Monkhorst1976}%
  \BibitemOpen
  \bibfield  {author} {\bibinfo {author} {\bibfnamefont {H.~J.}\ \bibnamefont
  {Monkhorst}}\ and\ \bibinfo {author} {\bibfnamefont {J.~D.}\ \bibnamefont
  {Pack}},\ }\href {\doibase 10.1103/PhysRevB.13.5188} {\bibfield  {journal}
  {\bibinfo  {journal} {Phys. Rev. B}\ }\textbf {\bibinfo {volume} {13}},\
  \bibinfo {pages} {5188} (\bibinfo {year} {1976})}\BibitemShut {NoStop}%
\bibitem [{\citenamefont {Kresse}\ and\ \citenamefont
  {Furthm{\"{u}}ller}(1996)}]{Kresse1996a}%
  \BibitemOpen
  \bibfield  {author} {\bibinfo {author} {\bibfnamefont {G.}~\bibnamefont
  {Kresse}}\ and\ \bibinfo {author} {\bibfnamefont {J.}~\bibnamefont
  {Furthm{\"{u}}ller}},\ }\href {\doibase 10.1103/PhysRevB.54.11169} {\bibfield
   {journal} {\bibinfo  {journal} {Phys. Rev. B}\ }\textbf {\bibinfo {volume}
  {54}},\ \bibinfo {pages} {11169} (\bibinfo {year} {1996})}\BibitemShut
  {NoStop}%
\bibitem [{\citenamefont {Wu}\ and\ \citenamefont {Zeng}(2016)}]{Wu2016a}%
  \BibitemOpen
  \bibfield  {author} {\bibinfo {author} {\bibfnamefont {M.}~\bibnamefont
  {Wu}}\ and\ \bibinfo {author} {\bibfnamefont {X.~C.}\ \bibnamefont {Zeng}},\
  }\href {\doibase 10.1021/acs.nanolett.6b00726} {\bibfield  {journal}
  {\bibinfo  {journal} {Nano Lett.}\ }\textbf {\bibinfo {volume} {16}},\
  \bibinfo {pages} {3236} (\bibinfo {year} {2016})}\BibitemShut {NoStop}%
\bibitem [{\citenamefont {Gomes}\ and\ \citenamefont
  {Carvalho}(2015)}]{Gomes2015}%
  \BibitemOpen
  \bibfield  {author} {\bibinfo {author} {\bibfnamefont {L.~C.}\ \bibnamefont
  {Gomes}}\ and\ \bibinfo {author} {\bibfnamefont {A.}~\bibnamefont
  {Carvalho}},\ }\href {\doibase 10.1103/PhysRevB.92.085406} {\bibfield
  {journal} {\bibinfo  {journal} {Phys. Rev. B}\ }\textbf {\bibinfo {volume}
  {92}},\ \bibinfo {pages} {085406} (\bibinfo {year} {2015})}\BibitemShut
  {NoStop}%
\bibitem [{\citenamefont {Guo}\ \emph {et~al.}(2015)\citenamefont {Guo},
  \citenamefont {Wang}, \citenamefont {Kuang},\ and\ \citenamefont
  {Huang}}]{Guo2015}%
  \BibitemOpen
  \bibfield  {author} {\bibinfo {author} {\bibfnamefont {R.}~\bibnamefont
  {Guo}}, \bibinfo {author} {\bibfnamefont {X.}~\bibnamefont {Wang}}, \bibinfo
  {author} {\bibfnamefont {Y.}~\bibnamefont {Kuang}}, \ and\ \bibinfo {author}
  {\bibfnamefont {B.}~\bibnamefont {Huang}},\ }\href {\doibase
  10.1103/PhysRevB.92.115202} {\bibfield  {journal} {\bibinfo  {journal} {Phys.
  Rev. B}\ }\textbf {\bibinfo {volume} {92}},\ \bibinfo {pages} {115202}
  (\bibinfo {year} {2015})}\BibitemShut {NoStop}%
\bibitem [{\citenamefont {Ding}\ \emph {et~al.}(2015)\citenamefont {Ding},
  \citenamefont {Gao},\ and\ \citenamefont {Yao}}]{Ding2015a}%
  \BibitemOpen
  \bibfield  {author} {\bibinfo {author} {\bibfnamefont {G.}~\bibnamefont
  {Ding}}, \bibinfo {author} {\bibfnamefont {G.}~\bibnamefont {Gao}}, \ and\
  \bibinfo {author} {\bibfnamefont {K.}~\bibnamefont {Yao}},\ }\href {\doibase
  10.1038/srep09567} {\bibfield  {journal} {\bibinfo  {journal} {Sci. Rep.}\
  }\textbf {\bibinfo {volume} {5}},\ \bibinfo {pages} {9567} (\bibinfo {year}
  {2015})}\BibitemShut {NoStop}%
\bibitem [{\citenamefont {Li}\ \emph {et~al.}(2017)\citenamefont {Li},
  \citenamefont {Bauers}, \citenamefont {Poudel}, \citenamefont {Hamann},
  \citenamefont {Wang}, \citenamefont {Choi}, \citenamefont {Esfarjani},
  \citenamefont {Shi}, \citenamefont {Johnson},\ and\ \citenamefont
  {Cronin}}]{Li2017}%
  \BibitemOpen
  \bibfield  {author} {\bibinfo {author} {\bibfnamefont {Z.}~\bibnamefont
  {Li}}, \bibinfo {author} {\bibfnamefont {S.~R.}\ \bibnamefont {Bauers}},
  \bibinfo {author} {\bibfnamefont {N.}~\bibnamefont {Poudel}}, \bibinfo
  {author} {\bibfnamefont {D.}~\bibnamefont {Hamann}}, \bibinfo {author}
  {\bibfnamefont {X.}~\bibnamefont {Wang}}, \bibinfo {author} {\bibfnamefont
  {D.~S.}\ \bibnamefont {Choi}}, \bibinfo {author} {\bibfnamefont
  {K.}~\bibnamefont {Esfarjani}}, \bibinfo {author} {\bibfnamefont
  {L.}~\bibnamefont {Shi}}, \bibinfo {author} {\bibfnamefont {D.~C.}\
  \bibnamefont {Johnson}}, \ and\ \bibinfo {author} {\bibfnamefont {S.~B.}\
  \bibnamefont {Cronin}},\ }\href {\doibase 10.1021/acs.nanolett.6b05402}
  {\bibfield  {journal} {\bibinfo  {journal} {Nano Lett.}\ }\textbf {\bibinfo
  {volume} {17}},\ \bibinfo {pages} {1978} (\bibinfo {year}
  {2017})}\BibitemShut {NoStop}%
\bibitem [{\citenamefont {Mahan}(2000)}]{Mahan2000}%
  \BibitemOpen
  \bibfield  {author} {\bibinfo {author} {\bibfnamefont {G.~D.}\ \bibnamefont
  {Mahan}},\ }\href {\doibase 10.1007/978-1-4757-5714-9} {\emph {\bibinfo
  {title} {{Many-Particle Physics}}}}\ (\bibinfo  {publisher} {Springer US},\
  \bibinfo {address} {Boston, MA},\ \bibinfo {year} {2000})\BibitemShut
  {NoStop}%
\bibitem [{\citenamefont {Hwang}\ and\ \citenamefont {{Das
  Sarma}}(2008)}]{Hwang2008}%
  \BibitemOpen
  \bibfield  {author} {\bibinfo {author} {\bibfnamefont {E.~H.}\ \bibnamefont
  {Hwang}}\ and\ \bibinfo {author} {\bibfnamefont {S.}~\bibnamefont {{Das
  Sarma}}},\ }\href {\doibase 10.1103/PhysRevB.77.115449} {\bibfield  {journal}
  {\bibinfo  {journal} {Phys. Rev. B}\ }\textbf {\bibinfo {volume} {77}},\
  \bibinfo {pages} {115449} (\bibinfo {year} {2008})}\BibitemShut {NoStop}%
\bibitem [{\citenamefont {Kaasbjerg}\ \emph {et~al.}(2012)\citenamefont
  {Kaasbjerg}, \citenamefont {Thygesen},\ and\ \citenamefont
  {Jacobsen}}]{Kaasbjerg2012}%
  \BibitemOpen
  \bibfield  {author} {\bibinfo {author} {\bibfnamefont {K.}~\bibnamefont
  {Kaasbjerg}}, \bibinfo {author} {\bibfnamefont {K.~S.}\ \bibnamefont
  {Thygesen}}, \ and\ \bibinfo {author} {\bibfnamefont {K.~W.}\ \bibnamefont
  {Jacobsen}},\ }\href {\doibase 10.1103/PhysRevB.85.115317} {\bibfield
  {journal} {\bibinfo  {journal} {Phys. Rev. B}\ }\textbf {\bibinfo {volume}
  {85}},\ \bibinfo {pages} {115317} (\bibinfo {year} {2012})}\BibitemShut
  {NoStop}%
\bibitem [{\citenamefont {Zhao}\ \emph
  {et~al.}(2016{\natexlab{b}})\citenamefont {Zhao}, \citenamefont {Shi},
  \citenamefont {Xi}, \citenamefont {Wang},\ and\ \citenamefont
  {Shuai}}]{Zhao2016}%
  \BibitemOpen
  \bibfield  {author} {\bibinfo {author} {\bibfnamefont {T.}~\bibnamefont
  {Zhao}}, \bibinfo {author} {\bibfnamefont {W.}~\bibnamefont {Shi}}, \bibinfo
  {author} {\bibfnamefont {J.}~\bibnamefont {Xi}}, \bibinfo {author}
  {\bibfnamefont {D.}~\bibnamefont {Wang}}, \ and\ \bibinfo {author}
  {\bibfnamefont {Z.}~\bibnamefont {Shuai}},\ }\href {\doibase
  10.1038/srep19968} {\bibfield  {journal} {\bibinfo  {journal} {Sci. Rep.}\
  }\textbf {\bibinfo {volume} {7}},\ \bibinfo {pages} {19968} (\bibinfo {year}
  {2016}{\natexlab{b}})}\BibitemShut {NoStop}%
\bibitem [{\citenamefont {Liu}\ \emph {et~al.}(2017)\citenamefont {Liu},
  \citenamefont {Zhou}, \citenamefont {Liao}, \citenamefont {Singh},\ and\
  \citenamefont {Chen}}]{Liu2017}%
  \BibitemOpen
  \bibfield  {author} {\bibinfo {author} {\bibfnamefont {T.~H.}\ \bibnamefont
  {Liu}}, \bibinfo {author} {\bibfnamefont {J.}~\bibnamefont {Zhou}}, \bibinfo
  {author} {\bibfnamefont {B.}~\bibnamefont {Liao}}, \bibinfo {author}
  {\bibfnamefont {D.~J.}\ \bibnamefont {Singh}}, \ and\ \bibinfo {author}
  {\bibfnamefont {G.}~\bibnamefont {Chen}},\ }\href {\doibase
  10.1103/PhysRevB.95.075206} {\bibfield  {journal} {\bibinfo  {journal} {Phys.
  Rev. B}\ }\textbf {\bibinfo {volume} {95}},\ \bibinfo {pages} {075206}
  (\bibinfo {year} {2017})}\BibitemShut {NoStop}%
\bibitem [{\citenamefont {Nag}(1980)}]{Nag1980}%
  \BibitemOpen
  \bibfield  {author} {\bibinfo {author} {\bibfnamefont {B.~R.}\ \bibnamefont
  {Nag}},\ }\href {\doibase 10.1007/978-3-642-81416-7} {\emph {\bibinfo {title}
  {{Electron Transport in Compound Semiconductors}}}}\ (\bibinfo  {publisher}
  {Springer-Verlag Berlin Heidelberg},\ \bibinfo {address} {Berlin,
  Heidelberg},\ \bibinfo {year} {1980})\BibitemShut {NoStop}%
\bibitem [{\citenamefont {Attaccalite}\ \emph {et~al.}(2010)\citenamefont
  {Attaccalite}, \citenamefont {Wirtz}, \citenamefont {Lazzeri}, \citenamefont
  {Mauri},\ and\ \citenamefont {Rubio}}]{Attaccalite2010}%
  \BibitemOpen
  \bibfield  {author} {\bibinfo {author} {\bibfnamefont {C.}~\bibnamefont
  {Attaccalite}}, \bibinfo {author} {\bibfnamefont {L.}~\bibnamefont {Wirtz}},
  \bibinfo {author} {\bibfnamefont {M.}~\bibnamefont {Lazzeri}}, \bibinfo
  {author} {\bibfnamefont {F.}~\bibnamefont {Mauri}}, \ and\ \bibinfo {author}
  {\bibfnamefont {A.}~\bibnamefont {Rubio}},\ }\href {\doibase
  10.1021/nl9034626} {\bibfield  {journal} {\bibinfo  {journal} {Nano Lett.}\
  }\textbf {\bibinfo {volume} {10}},\ \bibinfo {pages} {1172} (\bibinfo {year}
  {2010})}\BibitemShut {NoStop}%
\bibitem [{ref()}]{reference2017}%
  \BibitemOpen
  \href@noop {} {\bibinfo  {journal} {We would notice that the EPC strength in
  Ref.~[\onlinecite{Ulstrup2012}] is dimensionless and is obtained according to
  the definition
  \cite{Hofmann2009}:~$\lambda(\varepsilon_{i\mathbf{k}})=\int_{0}^{\omega_{\rm
  max}}\left[{\alpha^2{\mathcal
  F}^E(\varepsilon_{i\mathbf{k},\omega})+\alpha^2{\mathcal
  F}^A(\varepsilon_{i\mathbf{k},\omega})}\right]d\omega/\omega$, where
  $\lambda(\varepsilon_{i\mathbf{k}})$ represents the EPC parameter,
  $\omega_{\rm max}$ is the maximum phonon frequency, $\alpha^2{\mathcal
  F}^E(\varepsilon_{i\mathbf{k},\omega})$ is the Eliashberg coupling function
  corresponding to the phonon emission process and $\alpha^2{\mathcal
  F}^A(\varepsilon_{i\mathbf{k},\omega})$ for phonon absorption process. The
  definition seems different with ours shown in Eq.~(1) due to the additional
  factor $\omega$ in denominator. However, one can remain safely regard that
  the EPC strength of pristine graphene compared to that of the 2D SnSe is
  small since the phonon energy in general is much less than 1 eV}\
  }\BibitemShut {NoStop}%
\bibitem [{\citenamefont {Ziman}(1960)}]{Ziman1960}%
  \BibitemOpen
\bibfield  {journal} {  }\bibfield  {author} {\bibinfo {author} {\bibfnamefont
  {J.~M.}\ \bibnamefont {Ziman}},\ }\href {\doibase
  10.1016/0022-3697(60)90260-2} {\emph {\bibinfo {title} {{Electrons and
  phonons: The theory of transport phenomena in solids}}}}\ (\bibinfo
  {publisher} {Oxford University Press},\ \bibinfo {address} {New York},\
  \bibinfo {year} {1960})\BibitemShut {NoStop}%
\bibitem [{\citenamefont {Onsager}(1931{\natexlab{a}})}]{Onsager1931}%
  \BibitemOpen
  \bibfield  {author} {\bibinfo {author} {\bibfnamefont {L.}~\bibnamefont
  {Onsager}},\ }\href {\doibase 10.1103/PhysRev.37.405} {\bibfield  {journal}
  {\bibinfo  {journal} {Phys. Rev.}\ }\textbf {\bibinfo {volume} {37}},\
  \bibinfo {pages} {405} (\bibinfo {year} {1931}{\natexlab{a}})}\BibitemShut
  {NoStop}%
\bibitem [{\citenamefont {Onsager}(1931{\natexlab{b}})}]{Onsager1931a}%
  \BibitemOpen
  \bibfield  {author} {\bibinfo {author} {\bibfnamefont {L.}~\bibnamefont
  {Onsager}},\ }\href {\doibase 10.1103/PhysRev.38.2265} {\bibfield  {journal}
  {\bibinfo  {journal} {Phys. Rev.}\ }\textbf {\bibinfo {volume} {38}},\
  \bibinfo {pages} {2265} (\bibinfo {year} {1931}{\natexlab{b}})}\BibitemShut
  {NoStop}%
\bibitem [{\citenamefont {Jiang}\ and\ \citenamefont {Wang}(2011)}]{Jiang2011}%
  \BibitemOpen
  \bibfield  {author} {\bibinfo {author} {\bibfnamefont {J.~W.}\ \bibnamefont
  {Jiang}}\ and\ \bibinfo {author} {\bibfnamefont {J.-S.}\ \bibnamefont
  {Wang}},\ }\href {\doibase 10.1063/1.3671069} {\bibfield  {journal} {\bibinfo
   {journal} {J. Appl. Phys.}\ }\textbf {\bibinfo {volume} {110}},\ \bibinfo
  {pages} {124319} (\bibinfo {year} {2011})}\BibitemShut {NoStop}%
\bibitem [{\citenamefont {L{\"{u}}}\ \emph {et~al.}(2016)\citenamefont
  {L{\"{u}}}, \citenamefont {Wang}, \citenamefont {Hedeg{\aa}rd},\ and\
  \citenamefont {Brandbyge}}]{Lu2016}%
  \BibitemOpen
  \bibfield  {author} {\bibinfo {author} {\bibfnamefont {J.~T.}\ \bibnamefont
  {L{\"{u}}}}, \bibinfo {author} {\bibfnamefont {J.-S.}\ \bibnamefont {Wang}},
  \bibinfo {author} {\bibfnamefont {P.}~\bibnamefont {Hedeg{\aa}rd}}, \ and\
  \bibinfo {author} {\bibfnamefont {M.}~\bibnamefont {Brandbyge}},\ }\href
  {\doibase 10.1103/PhysRevB.93.205404} {\bibfield  {journal} {\bibinfo
  {journal} {Phys. Rev. B}\ }\textbf {\bibinfo {volume} {93}},\ \bibinfo
  {pages} {205404} (\bibinfo {year} {2016})}\BibitemShut {NoStop}%
\bibitem [{\citenamefont {Yang}\ \emph {et~al.}(2016)\citenamefont {Yang},
  \citenamefont {Perfetto}, \citenamefont {Kurth}, \citenamefont {Stefanucci},\
  and\ \citenamefont {D'Agosta}}]{Yang2016b}%
  \BibitemOpen
  \bibfield  {author} {\bibinfo {author} {\bibfnamefont {K.}~\bibnamefont
  {Yang}}, \bibinfo {author} {\bibfnamefont {E.}~\bibnamefont {Perfetto}},
  \bibinfo {author} {\bibfnamefont {S.}~\bibnamefont {Kurth}}, \bibinfo
  {author} {\bibfnamefont {G.}~\bibnamefont {Stefanucci}}, \ and\ \bibinfo
  {author} {\bibfnamefont {R.}~\bibnamefont {D'Agosta}},\ }\href {\doibase
  10.1103/PhysRevB.94.081410} {\bibfield  {journal} {\bibinfo  {journal} {Phys.
  Rev. B}\ }\textbf {\bibinfo {volume} {94}},\ \bibinfo {pages} {081410}
  (\bibinfo {year} {2016})}\BibitemShut {NoStop}%
\bibitem [{\citenamefont {Hofmann}\ \emph {et~al.}(2009)\citenamefont
  {Hofmann}, \citenamefont {Sklyadneva}, \citenamefont {Rienks},\ and\
  \citenamefont {Chulkov}}]{Hofmann2009}%
  \BibitemOpen
  \bibfield  {author} {\bibinfo {author} {\bibfnamefont {P.}~\bibnamefont
  {Hofmann}}, \bibinfo {author} {\bibfnamefont {I.~Y.}\ \bibnamefont
  {Sklyadneva}}, \bibinfo {author} {\bibfnamefont {E.~D.~L.}\ \bibnamefont
  {Rienks}}, \ and\ \bibinfo {author} {\bibfnamefont {E.~V.}\ \bibnamefont
  {Chulkov}},\ }\href {\doibase 10.1088/1367-2630/11/12/125005} {\bibfield
  {journal} {\bibinfo  {journal} {New J. Phys.}\ }\textbf {\bibinfo {volume}
  {11}},\ \bibinfo {pages} {125005} (\bibinfo {year} {2009})}\BibitemShut
  {NoStop}%
\end{thebibliography}%
\clearpage
\end{document}